\documentclass[11pt,a4paper,final]{iopart}

\usepackage{iopams}
\usepackage{cite}
\usepackage{graphicx}
\usepackage{epsfig}
\usepackage[breaklinks=true,colorlinks=true,linkcolor=blue,urlcolor=blue,citecolor=blue]{hyperref}
\usepackage{ae}
\usepackage[T1]{fontenc}
\usepackage[latin1]{inputenc}
\usepackage[english]{babel}
\usepackage{amssymb}
\usepackage{makeidx}
\usepackage{latexsym}

\usepackage[all]{xy}
\usepackage{graphicx}
\usepackage{fancyhdr}
\usepackage{theorem}
\usepackage{bbm}
\usepackage{latexsym}
\usepackage{psfrag}
\usepackage{slashed}
\usepackage{floatflt}
\usepackage{pstricks}
\usepackage{boxedminipage}

\usepackage[margin=2.8cm]{geometry}

\newcommand\gs{g_{\rm s}}
\newcommand\as{\alpha_{\rm s}}

\newcommand{\lp}{\left}
\newcommand{\rp}{\right}

\newcommand{\angIn}[1]{\left|\:#1\rangle\right.}
\renewcommand{\d}{\mathrm{d}}

\newcommand{\ord}{\mathcal{O}}
\newcommand{\MSb}{\overline{\rm MS}}

\newcommand{\TiTj}{{\bf T}_i \cdot {\bf T}_j}
\newcommand{\zcut}{z_{\rm cut}}

\newcommand{\cf}{C_{F}}
\newcommand{\ca}{C_{A}}
\newcommand{\nf}{n_{F}}
\renewcommand{\e}{\varepsilon}
\newcommand{\C}[2]{C^{(#2)}_{#1}}
\newcommand{\ea}{{\C{1}{\alpha}}}

\def\comix{\textsc{Comix} }
\def\cascade{\textsc{Cascade} }
\def\hej{\textsc{Hej} }
\def\herwig{\textsc{Herwig} }
\def\madgraph{\textsc{Madgraph} }
\def\caesar{\textsc{Caesar} }

\begin{document}
\begin{flushright}
MIT-CTP 4672 \\
MPP-2015-102
\end{flushright}

\topical{QCD resummation for hadronic final states}

\author{Gionata Luisoni}
\address{Max Plack Institute for Physics, Munich, Germany}
\ead{luisonig@mpp.mpg.de}

\author{Simone Marzani}
\address{Center for Theoretical Physics, Massachusetts Institute of Technology, \\77 Massachusetts Avenue, Cambridge, MA 02139, USA}
\ead{smarzani@mit.edu}

\begin{abstract}
We review the basic concepts of all-order calculations in Quantum Chromodynamics (QCD) and their application to collider phenomenology. We start by discussing the factorization properties of QCD amplitudes and cross-sections in the soft and collinear limits and their resulting all-order exponentiation. We then discuss several applications of this formalism to observables which are of great interest at particle colliders. In this context, we describe the all-order resummation of event-shape distributions, as well as observables that probe the internal structure of hadronic jets.
\end{abstract}

\section{Introduction}\label{sec:intro}

The CERN Large Hadron Collider (LHC) has recently resumed its operations after the first long shutdown. Run I of the LHC was extremely successful, with the milestone discovery of the long-sought-after Higgs boson by the ATLAS~\cite{Aad:2012tfa} and CMS~\cite{Chatrchyan:2012ufa} collaborations. Moreover, many other aspects of the Standard Model have been probed in a previously unexplored energy regime. As a result, after LHC Run I, the Standard Model appears now as a fully consistent and highly-successful theory of particle physics. However, very strong evidence of its incomplete nature already exists, above all the fact that no Standard Model particle appears to be a good candidate for dark matter.

This situation results into a two-fold task for LHC Run II. On the one hand, precise measurements of the Higgs boson properties are necessary, in order to verify whether it is fully responsible for electroweak symmetry breaking as predicted by the Brout-Englert-Higgs mechanism~\cite{Englert:1964et,Higgs:1964ia,Higgs:1964pj,Higgs:1966ev}. On the other hand, searches for new particles and possible inconsistencies in Standard Model predictions are going to be pushed to a new energy frontier. In order for these tasks to be successful and to fully exploit the LHC physics potential, new and more accurate theoretical and experimental tools have been, and are being, developed. The aim of this topical review is to discuss one of these theoretical tools, namely all-order calculations in perturbative Quantum Chromodynamics (QCD), the theory of strong interactions.

The LHC collides protons, which are strongly interacting, and further strongly-interacting particles are abundantly produced in every such collision. Therefore, careful studies of QCD radiation in Higgs production and new physics processes can be exploited in order to better understand their properties.
Moreover, the possibility of making discoveries at the LHC depends on our ability to separate new and rare phenomena from an overwhelming background, which is often several orders of magnitude bigger than the signal. This background consists of Standard Model processes and its dominant component comes from strong interactions.
It follows that precision QCD is mandatory at the LHC.
Furthermore, we need an accurate understanding of physics in the presence of disparate energy scales, which range from the unprecedentedly large colliding energy, through the electroweak scale, all the way down to hadron masses.
The appearance of multiple scales renders perturbative QCD calculations unreliable at any finite order. Therefore, an all-order re-organization of the perturbative expansion is necessary in order obtain reliable theoretical predictions.

Finally, despite being often taken for granted, it is far from trivial that precise perturbative QCD predictions can be compared to observables which are measured in terms of final-state hadrons. As we are going to discuss in the following, there exist special classes of observables for which corrections due to the hadronization process scale with inverse powers of the energy involved in the hard scattering and can therefore be considered reasonably under control at sufficiently high energies. Moreover, since it is not possible to compute these corrections from first principles, the use accurate parton-level predictions when comparing to data, allows us to constrain different ways of modeling the hadronization process. We will briefly come back to this in section~\ref{sec:event_shapes}.

In this topical review we discuss the basic concepts behind all-order calculations in QCD. More than one technique to perform such computations has been developed, and here we focus on analyses which directly make use of QCD matrix elements in the relevant soft and collinear limits. Analogous results can also be obtained using the methods of Soft-Collinear Effective Theory (SCET)~\cite{Bauer:2000ew,Bauer:2000yr,Bauer:2001ct,Bauer:2001yt,Beneke:2002ph,Beneke:2002ni,Hill:2002vw} (see Ref.~\cite{Becher:2014oda} for a recent, extensive, review). Moreover, recent studies have discussed the equivalence of the two methods~\cite{Sterman:2013nya,Almeida:2014uva,Bonvini:2014qga,Bonvini:2013td,Bonvini:2012az,Bonvini:2014tea}.

Even if we limit ourselves to resummation in direct QCD, where factorization is established for amplitudes and cross-sections in the soft or collinear limit, we still encounter different ways of obtaining all-order results. Generally speaking, we think one can identify an American school and a European (including Russia) one, although clear distinctions are difficult and sometimes misleading. The former, typically describes resummation by introducing non-local correlation operators, such as Wilson lines, and exploits their renormalization group evolution. The latter instead resorts to a more iterative procedure, directly identifying factorization and exponentiation properties of QCD matrix elements and cross-sections. In our presentation we will mostly follow this second approach.

In the first part of this review, we are going to discuss basic properties of gauge-theory amplitudes in the soft and collinear limits, specifically Quantum Electrodynamics (QED) and QCD, starting at one loop in section~\ref{sec:ir_divergences} and then moving to an all-order analysis in section~\ref{sec:resummation}.
In the second part of this review, we are going to focus on phenomenological applications. We will start with an analysis of event-shape variables in section~\ref{sec:event_shapes}, discussing thrust in some detail. We will then generalize our discussion to non-global observables in section~\ref{sec:non_global}. We will finish with a selection of more advanced topics in jet physics and jet substructure in section~\ref{sec:hadron_jets}, and a discussion about the limitations of factorization theorems upon which resummation is founded in section~\ref{sec:factorization}, before concluding.

For each topic, we are going to describe in some detail results that are well established in the literature, providing an extensive list of references. We are also going to provide short summaries of recent developments. We hope this way to stimulate the curiosity of both beginner and expert Readers.

\section{Infrared divergences} \label{sec:ir_divergences}

In order to better understand the origin of the corrections that we wish to resum, we start our discussion by making some considerations about the properties of scattering amplitudes in the soft and collinear limits.
In this regime, matrix elements involving massless particles may exhibit divergences. More specifically, the presence of soft divergences is related to the emission or exchange of particles with vanishing four-momentum, and it is associated with the presence of massless vector bosons. Such divergences occur even when the matter particles are massive. Collinear divergences are instead related to the splitting of particles at small angles, and are strictly present only when all involved particles are massless. Although these statements are rather general, in the following we will focus on QCD with some digression to QED, where several aspects can be simplified because of its Abelian nature.

An important point to make is that the strong coupling grows in the infrared and therefore perturbative QCD loses its predictive power at long distances. However, even without worrying about non-perturbative contributions, we also have the issue of how to define asymptotic states in gauge theories (QCD and QED alike), because of the presence of soft bremsstrahlung.

Our strategy is to define measurable quantities that are not affected by infrared and collinear (IRC) divergences. The Bloch-Nordsieck~(BN)~\cite{Blochnordsieck} and Kinoshita-Lee-Nauenberg~(KLN)~\cite{Kinoshita,Lee} theorems state that observable transition probabilities are free of IRC singularities. In our analysis we are going to consider \emph{safe} observables, i.e.\ measurable quantities that do not spoil the above theorems. We will come back to a more precise definition of IRC safety in sections~\ref{sec:qcd_soft_fact} and \ref{sec:event_shapes}. It is worth pointing out that from an experimental viewpoint, the finite resolution of the detectors acts as a regulator, thus preventing the occurrence of actual singularities. However, this in turn would be reflected on a possibly strong dependence of theoretical predictions on the detector resolution parameters, which we wish to avoid.

Even if we focus on IRC safe observables, from a practical viewpoint, when computing Feynman diagrams, we need a recipe in order to deal with potentially divergent intermediate contributions. We adopt the following
\begin{enumerate}
    \item{sum over all possible and indistinguishable processes that lead to the same final state configuration;}
    \item{compute cross-sections with an infrared regulator (e.g. in dimensional regularization with $d=4-2\varepsilon$ dimensions, where $\varepsilon<0$);}
    \item{select observables for which the $4$-dimensional limit is finite, i.e. that do not spoil BN and KLN theorems;}
    \item{interpret these results as perturbative estimate of the corresponding hadronic cross-section measured in experiments.}
\end{enumerate}

Thus, if we limit ourselves to the aforementioned class of \emph{safe} observables, perturbation theory will provide us with a finite result. However, as we shall discuss below, the cancellation of IRC singularities can leave behind a finite, but potentially large contribution, making the perturbative expansion unreliable at any finite order.
Our task is to identify these problematic IRC contributions and to re-organize the perturbative expansion in such a way that they are accounted for to all orders.
Even though the structure of IRC singularities in QCD is universal, in this section we will consider a simple example to illustrate their structure, namely the cross-section for electron-positron  ($e^+e^-$) annihilation into hadrons.

\subsection{Hadronic cross-section in $e^{+}e^{-}$ annihilation}

\begin{figure}
\begin{center}
\includegraphics[scale=1.1]{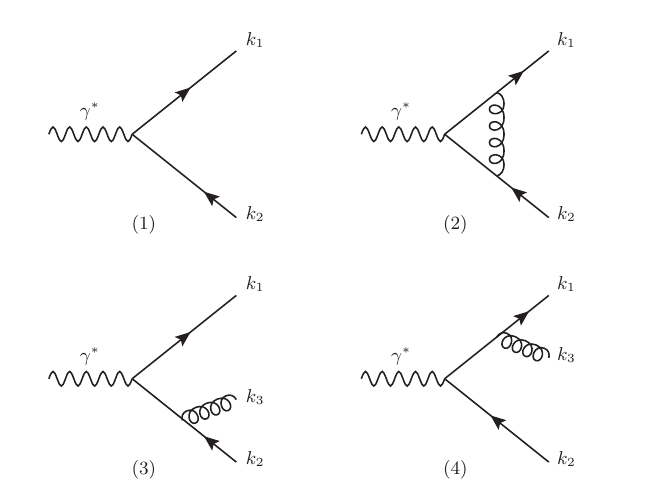}
\caption{Feynman diagrams contributing to the cross-section of $e^{+}e^{-}\to q\bar{q}g$.}\label{fig:sigma_nlo}
\end{center}
\end{figure}
The general expression for the cross-section $e^+e^- \to { \rm hadrons}$ can be written as
\begin{eqnarray}
\sigma_{\mathrm{had}}\lp( s\rp)
&=&\sigma_{\mathrm{had,0}}+\lp(\frac{\as}{2\pi}\rp)\sigma_{\mathrm{had,1}}+\lp(\frac{\as}{2\pi}\rp)^{2}\sigma_{\mathrm{had,2}}+\mathcal{O}\left(\as^3\right),
\end{eqnarray}
where $\as=\frac{g_s^2}{4 \pi}$ is the strong coupling constant.
The leading order (LO) contribution (Figure~\ref{fig:sigma_nlo}.1) can be cast as
\begin{equation}
\sigma_{\mathrm{had,0}}=\frac{4\pi\alpha}{3s}N_C\sum_{q}e_{q}^{2}\,,
\end{equation}
where we have introduced the fine structure constant $\alpha$, the quark electric charge $e_q$, the number of colors $N_C$ and the squared center-of-mass energy $s$.
At next-to-leading order (NLO) the real and virtual corrections have similar Feynman diagrams with different kinematics (Figure~\ref{fig:sigma_nlo}.3-\ref{fig:sigma_nlo}.4 and \ref{fig:sigma_nlo}.2 respectively). The real contribution is proportional to:
\begin{eqnarray}\label{eq:realee2jet}
\overline{\sum_{\mathrm{Spin}}}\lp|M_{3}\rp|^{2}&\propto&\lp(\frac{s_{13}}{s_{23}}+\frac{s_{23}}{s_{13}}+\frac{2s_{12}s_{123}}{s_{13}s_{23}}\rp)\,,
\end{eqnarray}
where $s_{ij}=\lp(k_i+k_j\rp)^{2}$ and $s_{ijk}=\lp(k_i+k_j+k_k\rp)^{2}$. The final state quark and anti-quark have momenta $k_1$ and $k_2$, whereas the gluon has momentum $k_3$. The sum in front of the squared 3-particle final state amplitude $M_{3}$ indicates the sum over the final state particle spins and the average over the initial once. The amplitude is singular in the limits:
\begin{equation}
s_{13}\to 0\,,\;s_{23}\to 0 \;\Leftrightarrow\;E_3\to 0 \,,\;\theta_{13}\to 0\,,\;\theta_{23}\to 0.
\end{equation}
Rewriting it in terms of energy fractions $x_i=2 k_i\cdot Q/Q^{2}$, where $Q=k_1+k_2+k_3$ is the center-of-mass energy, we can explicitly split the amplitude into a singular and a non-singular part
\begin{equation}
\fl
\quad \quad \quad
\overline{\sum_{\mathrm{Spin}}}\lp|M_{3}\rp|^{2}\propto\frac{x_1^{2}+x_2^{2}}{\lp(1-x_1\rp)\lp(1-x_2\rp)}
=\frac{\lp[1+\lp(1-x_3\rp)^{2}\rp]}{x_3}\lp(\frac{1}{1-x_1}+\frac{1}{1-x_2}\rp)-2.\label{eq:realee2jetsinglim}
\end{equation}
The above equation makes all singular limits explicit: the first term encodes the soft singularity and correspond to the Altarelli-Parisi splitting kernel
\begin{equation}
p_{gq}\lp(x\rp)=\frac{1+\lp(1-x\rp)^{2}}{x},
\end{equation}
which describes the probability of a collinear splitting. We can therefore write
\begin{equation}
\overline{\sum_{\mathrm{Spin}}}\lp|M_{3}\rp|^{2} \propto \frac{1}{1-x_1}p_{gq}\lp(x_3\rp)+\frac{1}{1-x_2}p_{gq}\lp(x_3\rp)-2.
\end{equation}
The first two terms describe the two possible collinear limits. The last term is instead finite. In order to evaluate the phase-space integral, we use dimensional regularization in $d=4-2\e$ dimensions.  The singularities now appear in terms of poles in $\e$\footnote{A step-by-step derivation of this result can be found e.g. in Ref.~\cite{Dissertori:2003pj}.}
\begin{equation}
\sigma_{\mathrm{had,1}}^{\mathrm{real}}=\sigma_{\mathrm{had,0}}\,\frac{\cf}{\Gamma\lp(1-\e\rp)}\lp(\frac{4\pi\mu^{2}}{s}\rp)^{\e}\lp(\frac{2}{\e^{2}}+\frac{3}{\e}-\pi^{2}+\frac{19}{2}+\mathcal{O}\lp(\e\rp)\rp).
\end{equation}
The virtual amplitude in Feynman gauge receives a contribution only from the vertex correction, which, once squared and integrated over phase space, gives
\begin{equation}
\sigma_{\mathrm{had,1}}^{\mathrm{virtual}}=\sigma_{\mathrm{had,0}}\,\frac{\cf}{\Gamma\lp(1-\e\rp)}\lp(\frac{4\pi\mu^{2}}{s}\rp)^{\e}\lp(-\frac{2}{\e^{2}}-\frac{3}{\e}+\pi^{2}-8+\mathcal{O}\lp(\e\rp)\rp).
\end{equation}
Therefore, the NLO contribution to the total cross-section is
\begin{equation}
\sigma_{\mathrm{had,1}}=\sigma_{\mathrm{had,1}}^{\mathrm{real}}+\sigma_{\mathrm{had,1}}^{\mathrm{virtual}}=\sigma_{\mathrm{had,0}}\,\frac{\cf}{\Gamma\lp(1-\e\rp)}\lp(\frac{4\pi\mu^{2}}{s}\rp)^{\e}\lp(\frac{3}{2}+\mathcal{O}\lp(\e\rp)\rp),
\end{equation}
and the total cross-section itself can be written as
\begin{equation}
\sigma_{\mathrm{had}}=\sigma_{\mathrm{had,0}}\lp(1+\frac{\as}{\pi}+\mathcal{O}\lp(\as^{2}\rp)\rp),
\end{equation}
which is finite. The real and the virtual contributions have the same singularity structure with opposite sign and the IRC poles in $\e$ cancel leaving a finite result, as stated by the KLN theorem. This cancellation occurs because in the soft and collinear limits both the real and the virtual amplitudes become proportional to the Born amplitude.
In what follows we are going to investigate this last point in more detail.

\subsection{Factorization in the soft limit}\label{sec:qcd_soft_fact}
In this section we study the factorization properties of real and virtual amplitudes in the soft limit.
We still focus on the above example and we start considering just one of the two diagrams (the one depicted in Fig.~\ref{fig:sigma_nlo}.4) contributing to the emission of a gluon from the quark anti-quark pair (for this reason the matrix element is primed). In order to stress the special role of the gluon in the soft limit, we are going to indicate its (soft) momentum by $q$, rather than $k_3$. We have
\begin{eqnarray}
M_{3}'&=&t_{1}^{a}\,\gs\,\mu^{\e}\bar{u}\lp(k_1\rp)\gamma^{\mu}\e_{\mu}\lp(q\rp)\frac{\slashed{q}+\slashed{k}_{1}}{\lp(q+k_1\rp)^{2}+i\e}\tilde{M}_{2}\nonumber\\
&\stackrel{q\to 0}{\longrightarrow}&t_{1}^{a}\,\gs\,\mu^{\e}\bar{u}\lp(k_1\rp)\gamma^{\mu}\e_{\mu}\lp(q\rp)\frac{\slashed{k}_{1}}{\lp(q+k_1\rp)^{2}+i\e}\tilde{M_{2}}\nonumber\\
&=&t_{1}^{a}\,\gs\,\mu^{\e}\frac{k_1^{\mu}}{k_1\cdot q}\e_{\mu}\lp(q\rp)M_{2}\,,
\end{eqnarray}
where the factor $k_1^{\mu}/(k_1\cdot q)$ is called eikonal factor and $t_{1}^{a}$ is the color matrix. We have also used fairly standard notation for the Dirac spinor $\bar u(k)$ and for the gluon polarization vector $\e_\mu(q)$. In the last step the Dirac spinor was absorbed in the 2-parton matrix element $M_2$ and therefore we dropped the tilde on it. For the full amplitude we find
\begin{equation}\label{eq:softfactorization}
\fl \quad \quad \quad
M_{3}\lp(p_1,p_2\to k_1,k_2,q\rp)\stackrel{q\to 0}{\longrightarrow} g_s \mu^{\e}J^{\mu}\lp(q\rp)\e_{\mu}\lp(q\rp)M_{2}\lp(p_1,p_2\to k_1,k_2\rp),
\end{equation}
where we have introduced the eikonal current
\begin{equation}\label{eq:eikonalcurrent}
J^{\mu}\lp(q\rp)=\sum_{n=1}^{2}t_n^{a}\frac{k_n^{\mu}}{k_n\cdot q}.
\end{equation}
It is important to note that the factorization does not depend on the internal structure of the amplitude. From the physical point of view, this reflects the fact that the large wavelength of the soft radiation cannot resolve the details of the short distance interactions. Squaring the amplitude, we obtain the following factorization property for the soft real emission:
\begin{eqnarray}\label{eq:eikonalfactorization}
\lp|M_3\rp|^{2}&\longrightarrow&\lp|M_2\rp|^{2}g_s^{2}\mu^{2\e}J^{\mu}\lp(q\rp)J^{\nu}\lp(q\rp)\lp(-g_{\mu\nu}\rp)\nonumber\\
&=&\lp|M_2\rp|^{2}g_s^{2}\mu^{2\e}\lp[-\sum_{m,n}t_{m}^{a}t_{n}^{a}\frac{k_m\cdot k_n}{(k_m\cdot q)(k_n\cdot q)}\rp]\nonumber\\
&=&\lp|M_2\rp|^{2}g_s^{2}\mu^{2\e}2\cf\frac{(k_1\cdot k_2)}{(k_1\cdot q)(k_2\cdot q)}\,.
\end{eqnarray}

The soft approximation can be applied also to the loop amplitude. In this limit we can in general neglect powers of the loop momentum $q$ in the numerator if $q^{\mu}\ll\sqrt{Q^{2}}$, furthermore in the denominator we can use the fact that $q^{2}\ll k_i\cdot q$. The loop correction to quark-antiquark pair production is therefore proportional to
\begin{eqnarray}
I&=&g_s^{2}\mu^{2\e}\cf(-i)\int\frac{\d^{d}q}{(2\pi)^{d}}\frac{\bar{u}\lp(k_1\rp)\gamma^{\mu}\lp(\slashed{k}_1+\slashed{q}\rp)\gamma^{\rho}\lp(\slashed{q}-\slashed{k}_2\rp)\gamma_{\mu}v\lp(k_2\rp)}{\lp[\lp(q+k_1\rp)^{2}+i\e\rp]\lp[\lp(q-k_2\rp)^{2}+i\e\rp]\lp[q^{2}+i\e\rp]}\nonumber\\
&\to&g_s^{2}\mu^{2\e}\cf(-i)\int\frac{\d^{d}q}{(2\pi)^{d}}\frac{\lp(k_1\cdot k_2\rp)\lp[\bar{u}\lp(k_1\rp)\gamma^{\rho}v\lp(k_2\rp)\rp]} {\lp[q\cdot k_1+i\e\rp]\lp[-q\cdot k_2+i\e\rp]\lp[q^{2}+i\e\rp]}.
\end{eqnarray}
The result in $d=4$ dimensions can be evaluated in the center-of-mass system of $k_1$ and $k_2$, with
\begin{equation}
\fl
k_1^{\mu}=E_1\lp(1,0,0,1\rp)\,,\quad k_2^{\mu}=E_2\lp(1,0,0,-1\rp)\,,\quad q^{\mu}=\lp(q_0,\vec{q}\rp)\,\textrm{ with }\, \vec{q}=\lp(\vec{q}_{\perp},q_z\rp)\,,
\end{equation}
where $\vec{q}_{\perp}$ is the vectorial transverse loop momentum and we define $q_{\perp}\equiv|\vec{q}_{\perp}|$. 

\noindent We thus obtain
\begin{equation}\label{eq:loopintsoft}
\fl \quad \quad
I=g_s^{2}\mu^{2\e}\cf(-i)\int\frac{\d^{3}q}{(2\pi)^{4}}\frac{2\:\d q_0 \lp[\bar{u}\lp(k_1\rp)\gamma^{\rho}v\lp(k_2\rp)\rp]}{\lp(q_0-q_z+i\e\rp)\lp(-q_0-q_z+i\e\rp)\lp(q_0^{2}-q_z^{2}-q_{\perp}^{2}+i\e\rp)}
\end{equation}
Eq.~(\ref{eq:loopintsoft}) has four poles in the complex $q_0$ plane at
\begin{eqnarray}
q_0=q_z-i\e,\qquad q_0=-q_z+i\e, \qquad q_0=\pm\lp(\lp|\vec{q\,}\rp|+i\e\rp)\,.
\end{eqnarray}
Closing the contour from below we find
\begin{equation}
\fl \quad \quad
I=g_s^{2}\mu^{2\e}\cf\lp[\bar{u}\lp(k_1\rp)\gamma^{\rho}v\lp(k_2\rp)\rp]\int\frac{\d^{3}q}{(2\pi)^{3}}\lp[\frac{-\lp(k_1\cdot k_2\rp)}{2\lp|\vec{q\,}\rp|\lp(k_1\cdot q\rp)\lp(k_2\cdot q\rp)}-\frac{1}{\lp(q_z-i\e\rp)\lp(q_\perp^{2}\rp)}\rp]\,,
\end{equation}
where the second integral is a pure phase
\begin{equation}
\fl \quad \quad
\int\frac{\d q_z\, \d^2 q_{\perp}}{\lp(2\pi\rp)^{3}}\frac{1}{\lp(q_z-i\e\rp)\lp(q_\perp^2 \rp)}=-\int \d q_z\frac{q_z+i\e}{q_z^{2}+\e^{2}}\int\frac{\d q_{\perp}}{\lp(2\pi\rp)^{2}}\frac{1}{q_\perp}=-\int\frac{\lp(i\pi\rp)}{\lp(2\pi\rp)^{2}}\frac{\d q_{\perp}}{q_\perp}\,.
\end{equation}
We are going to refer to the above contribution as Coulomb phase. We note that the above phase always cancels when considering physical cross-section in Abelian theories like QED. However, it can have a measurable effect in QCD cross-sections, in the presence of a high enough number of harder colored legs, as discussed in section~\ref{sec:caesar}.

Collecting real and virtual contributions together, we can compute the NLO distribution of an observable $v$ by introducing an appropriate observable-function $U_{i}\lp(\left\{k_i\right\}\rp)$:
\begin{eqnarray}
\sigma\lp(v\rp)&=&\frac{1}{2s}\int\d\Phi_2\lp|M_2\rp|^{2}U_2\lp(k_1,k_2\rp)+\nonumber\\
&&\frac{1}{2s}\int\d\Phi_2\lp|M_2\rp|^{2}\int\frac{\d^{3}q}{(2\pi)^{3}}\frac{1}{2\lp|q\rp|}2 g_s^{2}\cf\frac{\lp(k_1\cdot k_2\rp)}{\lp(k_1\cdot q\rp)\lp(k_2\cdot q\rp)}\nonumber\\
&&\hspace{3cm}\times\lp[U_{3}\lp(k_1,k_2,q\rp)-U_2\lp(k_1,k_2\rp)\rp]\,.
\end{eqnarray}
From the last equality we can derive the following important conclusions:
\begin{itemize}
\item[$\bullet$]{for a complete cancellation of the IRC contributions it is important that the observable is infrared and collinear safe, i.e. according to the definition of Ref.~\cite{Sterman:1977wj}, that it satisfies:
\begin{eqnarray}
\fl \quad \quad
U_{m+1}\lp(\ldots,k_i,k_j,\ldots\rp)&\longrightarrow& U_{m}\lp(\ldots,k_i+k_j,\ldots\rp) \qquad\mathrm{if}\;k_i\parallel k_j ,\\
\fl \quad \quad
U_{m+1}\lp(\ldots,k_i,\ldots\rp)&\longrightarrow& U_{m}\lp(\ldots,k_{i-1},k_{i+1},\ldots\rp) \quad\mathrm{if}\;k_i\to 0.
\end{eqnarray}
These limits have to hold not only for a single particle, but for an ensamble of partons becoming soft and/or collinear.
IRC safe properties of jet cross-sections and related variables, such as event shapes and energy correlation functions were first studied in Refs.~\cite{Sterman:1978bi,Sterman:1978bj,Sterman:1979uw}.
We note here that there exists a wealth of observables that are of great interest despite them being IRC unsafe. Generally speaking, these observables require the introduction of non-perturbative functions to describe their soft and/or collinear behavior. For example, lepton-hadron and hadron-hadron cross-sections are written as a momentum-fraction convolution of partonic cross-sections and parton distribution functions. Arbitrary collinear emissions change the value of the momentum fraction that enters the hard scattering, resulting in un-cancelled collinear singularities. Finite cross-sections are then obtained by a renormalization procedure of the parton densities.  Similar situations are also encountered in final-state evolution, if one is interested in measuring a particular type of hadron (see e.g.~\cite{Collins:1987pm}) or if the measurement only involves charged particles~\cite{Chang:2013rca,Chang:2013iba}.
Furthermore, we mention that recent work~\cite{Larkoski:2013paa, Larkoski:2014wba,Larkoski:2015lea} has introduced the concept of Sudakov safety, which enables to extend the reach of (resummed) perturbation theory beyond the IRC domain.}
\item[$\bullet$]{in the case of inclusive observables, for which $U_m\lp(k_1,\ldots,k_m\rp)=1$ for all $m$, the cancellation is complete. Consequently, the total cross-section remains unchanged by the emission of soft particles, as it should.}
\item[$\bullet$]{in case of an exclusive (but IRC safe) measurement, although the singularities cancel, the kinematic dependence of the observable can cause an unbalance between real and virtual contributions, which manifests itself with the appearance of potentially large logarithmic corrections to any orders in perturbation theory.}
\end{itemize}
These large logarithmic contributions spoil the perturbative expansion in the strong coupling and must be resummed to all orders in order to obtain reliable theoretical predictions for exclusive measurements.
In the following, we will discuss how the all-order behavior of QCD in the IRC regime can be systematically captured, enabling us to obtain perturbative predictions in kinematical regions where the fixed-order expansion in the strong coupling breaks down.

\section{A first look at all-order resummation: the coherent branching algorithm in the leading collinear approximation} \label{sec:resummation}
In this section we present a first all-order analysis of gauge theories (QED and QCD) based on the formalism of generating functionals~\cite{Dokshitzer:1991wu}. Our discussion follows the presentation given in the lecture series of Ref.~\cite{catani_notes}.
\subsection{Soft emission probability in QED}
We start by considering the factorization of a matrix element in the presence of a soft emission as derived in Eqs.~(\ref{eq:softfactorization}) and (\ref{eq:eikonalcurrent}), but this time we restrict ourselves to the case of an Abelian gauge theory like QED.
Labeling the hard momenta with $\left\{k_{i}\right\}$ and the momentum which becomes unresolved with $q$, and considering the possibility of emitting a photon by all the charged external legs, we can write
\begin{equation}\label{eq:softqedfactorization}
M_{m+1}\lp(\left\{k_{i}\right\},q\rp) \stackrel{q\to0}{\approx} g J^{\mu}\lp(q\rp)\varepsilon_{\mu}\lp(q\rp) M_{m}\lp(\left\{k_{i}\right\}\rp)\,,
\end{equation}
where $J^{\mu}$ is the eikonal current
\begin{equation}
J^{\mu}\lp(q\rp) = \sum_{i} e_{i} \frac{k_{i}^{\mu}}{k_i \cdot q}\,.
\end{equation}
The sign of the lepton charge $e_{i}$ is assigned depending on whether the momentum $k_i$ is incoming or outgoing. Because of charge conservation we have
\begin{equation}
\sum_{i}e_{i}=0.
\end{equation}
Squaring Eq. (\ref{eq:softqedfactorization}) and taking into account also the phase space we obtain an expression for the factorization of the cross-section
\begin{equation}
\d\sigma_{m+1}\lp(\left\{k_{i}\right\},q\rp) = \d\sigma_{m}\lp(\left\{k_{i}\right\}\rp)\times\d W_1\lp(q\rp)\,,
\end{equation}
where $\d W_1$ is the single photon emission probability. It can be written as
\begin{equation}\label{eq:single_photon_em_prob}
\d W_1\lp(q\rp)= \lp[\d q\rp] J^{\mu}\lp(q\rp)d_{\mu\nu} J^{\nu}\lp(q\rp) g^{2}\,,
\end{equation}
where $d_{\mu\nu}$ is the sum over the photon polarizations. Because of charge conservation only the term proportional to the metric tensor is different from zero when $d_{\mu\nu}$ is contracted with the eikonal currents. The result of this contraction can be cast into (we choose here an explicit reference frame for the momenta and define $\omega$ to be the energy of the soft photon)
\begin{equation}\label{eq:eiksquare}
J^{\mu}\lp(q\rp)J_{\mu}\lp(q\rp)=-\frac{2}{\omega^{2}}\sum_{i<j}e_{i}e_{j}\frac{1-\cos\theta_{ij}}{\lp(1-\cos\theta_{iq}\rp)\lp(1-\cos\theta_{jq}\rp)}\,.
\end{equation}
Note that Eq. (\ref{eq:eiksquare}) contains also the effects of the interference between emissions from different external lines. The physical consequences of this interference is important, we therefore analyze the last equation a bit further\footnote{The following considerations are derived in detail also in Chap 5.5 of the book by Ellis, Stirling and Webber~\cite{ellis}.}. We consider the term in the sum
\begin{equation}
W_{ij} \equiv \frac{1-\cos\theta_{ij}}{\lp(1-\cos\theta_{iq}\rp)\lp(1-\cos\theta_{jq}\rp)}\,,
\end{equation}
and split it into two terms
\begin{equation}\label{eq:angular_ordered_rad_func}
W_{ij}=W_{ij}^{\lp[i\rp]}+W_{ij}^{\lp[j\rp]}\,,
\end{equation}
where
\begin{equation}
W_{ij}^{\lp[i\rp]}=\frac{1}{2}\lp(W_{ij}+\frac{1}{1-\cos\theta_{ik}}-\frac{1}{1-\cos\theta_{jk}}\rp), \qquad W_{ij}^{\lp[j\rp]}=\lp.W_{ij}^{\lp[i\rp]}\rp|_{i\leftrightarrow j}\,.
\end{equation}
The two terms on the right hand side of Eq.~(\ref{eq:angular_ordered_rad_func}) have angular ordering properties. If we write the angular integration in terms of the polar and azimuthal angle with respect to the momentum $k_i$ and we integrate over the azimuth, we find
\begin{equation}\label{eq:ang_ord}
\int_{0}^{2\pi}\frac{\d\phi_{iq}}{2\pi}W_{ij}^{\lp[i\rp]}=\left\{
  \begin{array}{lr}
    \frac{1}{1-\cos\theta_{iq}} & \textrm{if }\theta_{iq}<\theta_{ij}, \\
    0 & \textrm{otherwise.}
  \end{array}
\right.
\end{equation}
This means that each of the two terms describe radiation which is confined in a cone. The physical interpretation behind this is that photons at larger angles are unable to resolve the pair of partons $(i,j)$ as separate charges. We consider now again the single photon emission probability of Eq.~(\ref{eq:single_photon_em_prob}) in the leading collinear approximation of $q$ being collinear to $k_j$. This means that we consider the angle $\theta_{jq}$ to be much smaller than any other relative angle. In this limit we can recast Eq.~(\ref{eq:eiksquare}) as
\begin{equation}
J^{2}\lp(q\rp)\stackrel{\theta_{jq}\to 0}{=}-\frac{1}{\omega^{2}}\lp[\frac{2e_j}{\lp(1-\cos\theta_{jq}\rp)}\sum_{i\neq j}e_{i}\frac{\lp(1-\cos\theta_{ij}\rp)}{\lp(1-\cos\theta_{iq}\rp)}+\mathcal{O}\lp(1\rp)\rp].
\end{equation}
The terms of $\mathcal{O}\lp(1\rp)$ are regular in the limit $\theta_{jq}\to 0$ and can be neglected whereas, in the considered limit, the sum can be approximated as
\begin{equation}
\sum_{i\neq j}e_{i}\frac{\lp(1-\cos\theta_{ij}\rp)}{\lp(1-\cos\theta_{ik}\rp)}\approx \sum_{i\neq j}e_{i} = -e_j.
\end{equation}
Writing the differential phase space element as
\begin{equation}
\lp[dq\rp]=\frac{\d^{3}q}{\lp(2\pi\rp)^{3}2\omega}=\frac{\omega^{2}\d\omega\,\d\phi\,\d\cos\theta}{\lp(2\pi\rp)^{3}2\omega},
\end{equation}
we can express the single photon emission probability as
\begin{equation}\label{eq:single_approx_photon_em_prob}
\d W_{1}\lp(q\rp)=\frac{\alpha}{\pi}\sum_{j}e_{j}^{2}\frac{\d\omega}{\omega}\frac{\d\theta_{jq}^{2}}{\theta_{jq}^{2}}\Theta\lp(\theta_{\max}-\theta_{jq}\rp)\,,
\end{equation}
where we have taken the limit of small angle $\theta$, as appropriate for the collinear region; $\theta_{\mathrm{max}}$ is of the order of the other angles $\theta_{ij}$. This means that we can approximate the single photon emission probability by considering a sum over radiation emitted by independent sources, where the destructive interference, which cancels the radiation at large angles, is approximated by the angular ordering constraint.

\subsection{Multiple soft emission and the generating functional method}

We can now generalize the emission of a single soft photon to multiple emissions from the same fermionic line. Since photons do not carry electric charge and since the emissions we consider are soft, they leave both the charge and the momentum of the emitting particle unchanged.
Therefore we can write the multiple emission probability as
\begin{equation}\label{eq:multiple_soft_emissions}
\d W_{n}\simeq\frac{1}{n!}\prod_{i=1}^{n}\d W_{1}\lp(q_i\rp)\,,
\end{equation}
where the prefactor is the symmetry factor for $n$ identical bosons.

The emission of multiple soft photons to all orders can conveniently be described by introducing a generating functional:
\begin{equation}\label{eq:generatin_func_qed}
\Phi^{\mathrm{real}}\lp[u\lp(q\rp)\rp] \equiv 1+\sum_{n=1}^{\infty}\int \d W_{n}\lp(q_1,\ldots,q_n\rp)u\lp(q_1\rp)\cdot\ldots\cdot u\lp(q_n\rp),
\end{equation}
where for each photon we introduced an arbitrary weight, which acts as phase-space constraint similarly to the observable function $U_m$ introduced in the previous section, but in the soft-collinear limit.
From Eq.~(\ref{eq:generatin_func_qed}) one can recover any emission probability by successive differentiation at $u=0$:
\begin{equation}
\d W\lp(q_1,\ldots,q_n\rp)=\lp.\frac{\delta\Phi}{\delta u\lp(q_1\rp)\cdot\ldots\cdot \delta u\lp(q_n\rp)}\rp|_{u=0}\,,
\end{equation}
Using Eq.~(\ref{eq:multiple_soft_emissions}) we can rewrite Eq.~(\ref{eq:generatin_func_qed}) as
\begin{eqnarray}
\fl \quad
\Phi^{\mathrm{real}}\lp[u\lp(q\rp)\rp]&=&1+\sum_{n=1}^{\infty}\frac{1}{n!}\prod_{i=1}^{n}\lp[\int\d W_1\lp(q_i\rp)u\lp(q_i\rp)\rp]
=1+\sum_{n=1}^{\infty}\frac{1}{n!}\lp[\int\d W_1\lp(q_i\rp)u\lp(q_i\rp)\rp]^{n}\nonumber\\
&=&\exp\lp\{\int\d W_1\lp(q\rp)u\lp(q\rp)\rp\}.
\end{eqnarray}
This equation gives the corrections due to real soft photon emission to a squared matrix element in QED. Form the last line we observe that it is simply given by exponentiating the lowest order contribution. This was derived for the first time in the 1960s in Ref.~\cite{Yennie:1961ad}. To find the total correction due to soft emission we also have to consider the virtual contribution. In the previous section we explicitly computed the loop contribution and found that the total (unconstrained) soft emission has a vanishing effect. We can therefore exploit this result by imposing:
\begin{equation}
\lp.\Phi\lp[u\lp(q\rp)\rp]\rp|_{u=1}=1\,.
\end{equation}
This is often referred to as the unitarity condition and allows us to correctly normalize $\Phi$, which can finally be written as
\begin{eqnarray}
\Phi\lp[u\lp(q\rp)\rp]&=&\frac{\Phi^{\mathrm{real}}\lp[u\lp(q\rp)\rp]}{\Phi^{\mathrm{real}}\lp[u\lp(q\rp)=1\rp]}\nonumber\\
&=&\exp\lp\{\int\d W_1\lp(q\rp)\lp[u(q)-1\rp]\Theta\lp(Q-\omega\rp)\Theta\lp(\omega\theta-Q_0\rp)\rp\}\,.
\end{eqnarray}
In the last equation we have introduced two constraints in terms of step-functions, the first one gives an upper limit on the energy considered in the soft approximation $Q$, which is typically of the order of the hard scale considered in the process. The second $\Theta$-function introduces an arbitrary lower cutoff $Q_0$ for the photon transverse momentum. The dependence on $Q_0$ does drop off  when IRC safe observables are considered. Note that by requiring that no real radiation is emitted, i.e. setting $u\lp(q\rp)=0$, we obtain the so-called the Sudakov form factor
\begin{equation}
\fl \quad \quad \quad
\Phi\lp[u=0\rp]\equiv \Delta\lp(Q,Q_0\rp)=\exp\lp\{-\int\d W_1\lp(q\rp)\Theta\lp(Q-\omega\rp)\Theta\lp(\omega\theta-Q_0\rp)\rp\},
\end{equation}
which describes the probability that no emission takes place and is fully determined by the virtual contributions. In the limit of vanishing lower cutoff $Q_0$, the exponentiated integral diverges and the Sudakov form factor vanishes as well. This means that the probability of not observing any radiation is zero. In other words, in any scattering process there must be some radiation, which can however be arbitrarily soft.

\begin{figure}
\centering
\includegraphics[scale=0.49]{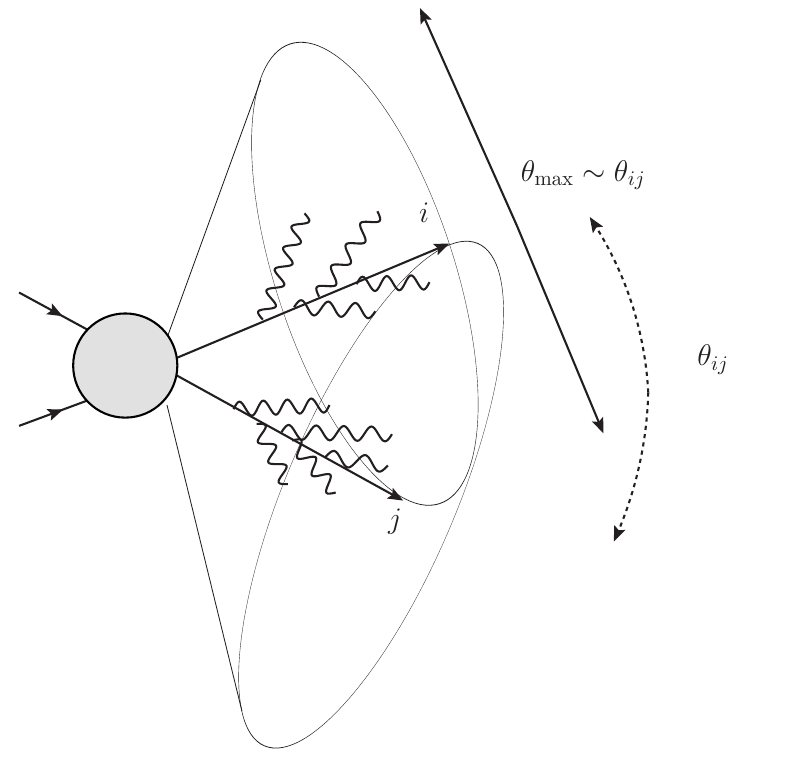}
\hfill
\includegraphics[scale=0.49]{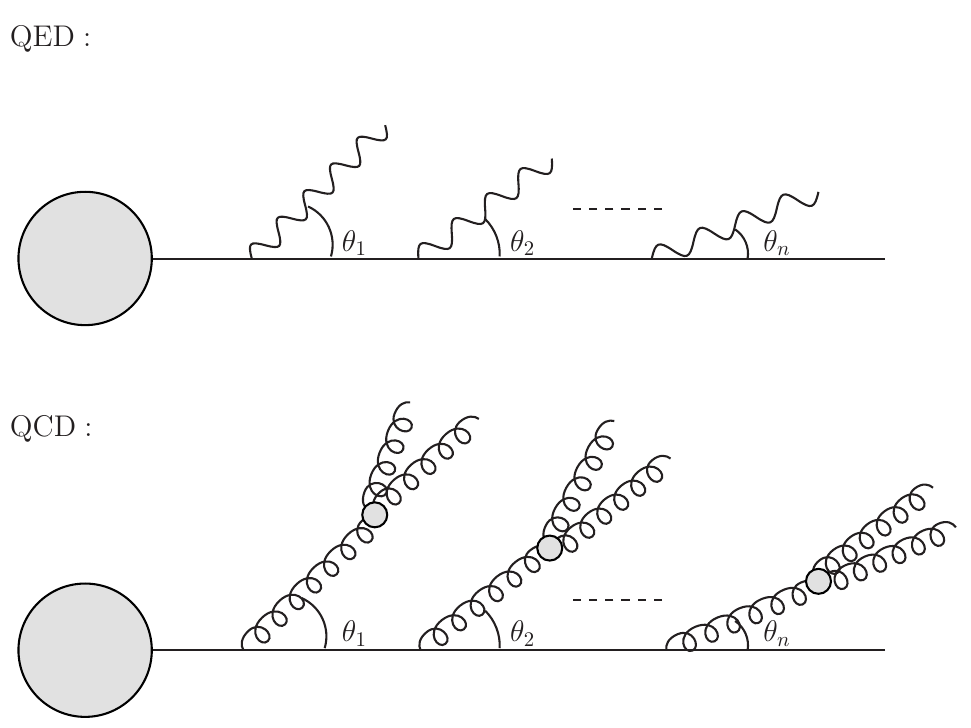}
\caption{Schematic representation of soft emissions in QED and QCD. On the left independent emission from different external legs in QED. On the right angular ordered emissions along an external leg in QED (top) and in QCD (bottom).}\label{fig:emission_in_qed_qcd}
\end{figure}

When considering multiple emissions from different charged fermions we have to remember the effect of interference, which can be approximated by the constraint on the angular ordering. For this case the generating functional can be written as
\begin{equation}
\Phi_{\lp\{k_1,\ldots,k_n\rp\}}\lp[u\lp(q\rp)\rp]=\prod_{i}\Phi_{k_i}\lp[Q,\theta_{\max};u\lp(q\rp)\rp]\,,
\end{equation}
where for each charged fermion external line we have
\begin{equation}
\fl \quad \quad \quad
\Phi_{k_i}\lp[Q,\theta_{\max};u\lp(q\rp)\rp]=\exp\lp\{\frac{\alpha}{\pi}\int_{0}^{Q}\frac{\d\omega}{\omega}\int_{0}^{\theta_{\max}}\frac{\d\theta^{2}}{\theta^{2}}\lp[u\lp(q\rp)-1\rp]\Theta\lp(\omega\theta-Q_0\rp)\rp\}\,.
\end{equation}
A schematic representation of the angular ordered emissions is given in Fig.~\ref{fig:emission_in_qed_qcd}.

\subsection{Generating functional in QCD} \label{sec:gen_func_QCD}
We now extend the generating functional method to QCD. The non-Abelian structure of the color charge leads to some complications in comparison with the derivation of the previous section. Charges have a matrix representation in color space and therefore the factorization in the soft limit is more involved. The example case used in section~\ref{sec:qcd_soft_fact} to derive the factorization in the soft limit has a very simple color structure. For processes with more colored particles Eqs.~(\ref{eq:eikonalcurrent}) and~(\ref{eq:eikonalfactorization}) have to be generalized. The eikonal current can be written as
\begin{equation}
J^{\mu}\lp(q\rp)=\sum_{i}{\bf T}_i\frac{k_i^{\mu}}{k_i\cdot q}\,,
\end{equation}
where the ${\bf T}_i$'s are abstract color operators (see for instance Refs.~\cite{Catani:1985xt,Catani:1996vz,Catani:1996jh,Dokshitzer:2005ig,Dokshitzer:2005ek}), which satisfy color conservation
\begin{equation} \label{eq:color_conservation}
\sum_i {\bf T}_i \angIn{M_m (\{k\} )}=0.
\end{equation}
In the last equation $\angIn{M_m (\{k\} )}$ is a vector in color space.
The square of an operator ${\bf T}_i$ gives the Casimir of the $SU(N_C)$ representation
\begin{equation}
{\bf T}_i^{2}=\left\{
  \begin{array}{ll}
    \cf & \textrm{if $i$ is a quark or anti-quark,} \\
    \ca & \textrm{if $i$ is a gluon,}
  \end{array}
\right .
\end{equation}
while, once a color basis is fixed, the color products $\TiTj$ are represented by matrices, which are, in general, non-diagonal. We note that in the large-$N_C$ limit, often considered, for instance, in Monte Carlo parton showers, the off-diagonal entries of these matrices vanish.

The factorization of an $(m+1)$-parton matrix element in the soft limit can thus be written as
\begin{equation}
\lp|M_{m+1}\rp|^{2}\longrightarrow -g_s^{2}\mu^{2\e}\sum_{i,j}\frac{(k_i\cdot k_j)}{(k_i\cdot q)(k_j\cdot q)}\lp\langle M_m|\TiTj|M_m\rp\rangle.
\end{equation}
An important consequence of the non-Abelian structure is that factorization is generally incomplete because of the presence of color matrices. Moreover, we also have to consider the fact that soft radiation can come from hard gluons and that, no matter how soft the emitted gluons are, they will always carry away charge. The exact pattern of soft gluon radiation is therefore more involved than in QED, and in order to keep this presentation simple, we limit ourselves to the leading collinear behavior, where we have remarkable simplifications. We shall come back to the more general case in section~\ref{sec:event_shapes}, where we consider the resummation of a specific class of observables.

In the leading collinear approximation, the probability for a single emission is very similar to the expression we have previously derived for QED, Eq.~(\ref{eq:single_approx_photon_em_prob}),
\begin{equation}
\d W_1\lp(q\rp)=\frac{\as}{\pi}\sum_{i}C_i \frac{\d\omega}{\omega}\frac{\d\theta_{iq}^{2}}{\theta_{iq}^{2}}\Theta\lp(\theta_{\max}-\theta_{iq}\rp)\,,
\end{equation}
where $C_i=C_F$ in case parton $i$ is a quark or an anti-quark and $C_i=C_A$ in case parton $i$ is a gluon. As in the case of photon emission interference effects can be captured by means of angular ordering. This is strictly true only at the logarithmic order we are considering. As we shall see later, wide-angle soft radiation, whose contribution is subleading, makes this pattern much more involved.

As already anticipated, when considering multiple gluon emission color correlations cannot be neglected. After the first splitting the total color charge will be shared among the two partons and further radiation can be emitted from either of them. This more complicated radiation pattern can be simplified with the help of color coherence. In fact, as derived in Eq.~(\ref{eq:ang_ord}) above, soft radiation cannot resolve the details of the interaction which happens at shorter distance and higher momentum scale. Therefore a soft gluon emitted at an angle $\theta_1$ will only ``see'' the total color charge of the radiation emitted at smaller angles $\theta<\theta_1$, which corresponds to the charge of the hardest parton involved~\cite{Ermolaev:1981cm,Mueller:1981ex,Bassetto:1982ma}. Iterating this argument we can describe multiple soft QCD radiation in the leading collinear approximation in a similar way to the QED case, remembering that all the radiation needs to be angular ordered. Furthermore, each emitted parton can act as a further emitter, as represented on the lower right picture of Fig.~\ref{fig:emission_in_qed_qcd}.
The final equation for the generating functional in leading collinear approximation in QCD can thus be written as
\begin{equation}\label{eq:phi_qcd}
\Phi_{\lp\{k_1,\ldots,k_n\rp\}}\lp[u(q)\rp]=\prod_{i}\Phi_{k_i}\lp[E_{k_i},\theta_{\max};u\lp(k\rp)\rp]\,,
\end{equation}
where however is it not possible to write a closed expression for $\Phi_{k_i}$, but only an iterative one~\cite{Dokshitzer:1982xr,Dokshitzer:1982ia,Catani:1984dp}
\begin{eqnarray}
\Phi_{k_i}\lp[E_{k_i},\theta_{\max}^{2};u\lp(q\rp)\rp]&=&\exp\lp\{\frac{\as C_{k_i}}{\pi}
\int_0^{E_{k_i}}\frac{\d\omega_{q}}{\omega_{q}}\int_0^{\theta_{\max}^{2}}\frac{\d\theta_{k_{i}q}^{2}}{\theta_{k_{i}q}^{2}}\rp.\nonumber\\
&&\lp.\times\lp(u\lp(q\rp)\Phi_q\lp[\omega_q,\theta_{k_iq};u\rp]-1\rp)\rp\}\,.\label{eq:generatin_func_qcd}
\end{eqnarray}
The first term in the second line describes further real radiation which is softer and occurs at a smaller angle, the second term encodes the virtual corrections and can be derived via a unitarity argument as done in the previous section.
Eqs. (\ref{eq:phi_qcd}) and (\ref{eq:generatin_func_qcd}) constitute the essence of the so called \emph{coherent branching algorithm}~\cite{Catani:1990rr,catani}. They are the starting point for constructing Monte Carlo parton shower algorithms~\cite{Marchesini:1983bm} such as \herwig~\cite{Marchesini:1991ch} and for the all-order resummation of the leading logarithms, which can be systematically improved, as we will discuss in the following.

From Eq.~(\ref{eq:generatin_func_qcd}), we can easily rederive the single emission probability, which reads
\begin{eqnarray}
\fl
\d W_{1}&=&\lp.\frac{\delta\Phi_{k}}{\delta u(k)}\rp|_{u\equiv0}
=\Phi_{k}\lp[E_k,\theta_{\max}^{2},0\rp]\lp[\frac{\as}{\pi}\cf\int_0^{E_k}\frac{\d\omega_q}{\omega_q}\int_{0}^{\theta_{\max}^{2}}\frac{\d\theta_{kq}^{2}}{\theta_{kq}^{2}}\Phi_{q}\lp(\omega_q,\theta_{kq},0\rp)\rp]\nonumber\\
\fl
&=&\Delta\lp(E_k,\theta_{\max}^{2},\theta_{kq}^{2}\rp)\lp[\frac{\as}{\pi}\cf\int_0^{E_k}\frac{\d\omega_q}{\omega_q}\int_{0}^{\theta_{\max}^{2}}\frac{\d\theta_{kq}^{2}}{\theta_{kq}^{2}}\Delta\lp(\omega_q,\theta_{kq},0\rp)\rp]\Delta\lp(E_k,\theta_{kq}^{2},0\rp)
\end{eqnarray}
In the last equation the first and the last Sudakov form factors describe the probability of evolving without branching of the mother particle before and after the emission. The emission probability is given by the term in square brackets, which includes the probability for the emitted particle to evolve without further branchings. The double emission probability can be obtained computing a further derivative. The recursiveness of Eq.~(\ref{eq:generatin_func_qcd}) will generate two terms in that case: one describing an angular ordered double emission from the same mother particle and another one describing a further splitting of the first emitted gluon.

The logarithmic accuracy of this description can be straightforwardly improved to include hard collinear radiation by considering the Altarelli-Parisi functions, rather than their soft limit, as splitting kernels, while the inclusion of soft emissions away from the collinear limit is more involved. Both of these issues will be discussed in the next section, where we are going to present all-order results for a wide class of observables, i.e.\ event shapes.

\section{Event shapes in $e^+e^-$ collisions} \label{sec:event_shapes}
Thus far we have discussed general properties of QCD matrix elements in the soft and collinear regime, showing how perturbative matrix elements factorize in those limits, leading to the possibility of all-order calculations in QCD.
We now turn our attention to the all-order behavior of specific classes of observables.

We would like to start with an example that although simple, contains most of the ingredients which form the building blocks of resummation technology.
A good candidate for this program is actually a fairly large class of observables called \emph{event shapes}, which aim to measure the energy flow of an event.
In order to avoid possible complications due to initial-state radiation and the role of parton distribution functions, we begin by considering event shapes in $e^+e^-$ annihilation.
Not only this simplifies our analysis, but it also has phenomenological value. Although event shapes can be defined and studied at hadron colliders, e.g.~\cite{Aad:2012np,Khachatryan:2014ika}, they are best measured at lepton colliders, such as LEP, where they are used to determine the strong coupling constant, e.g.~\cite{Abbiendi:2004qz,Dissertori:2007xa,Bethke:2008hf,Becher:2008cf,Abbate:2010xh,OPAL:2011aa,Gehrmann:2012sc}. The absence of some non-perturbative phenomena, such as the underlying event and pile-up, makes event shapes at lepton colliders an invaluable tool to study the energy-momentum flow of strongly interacting final states.

We are going to consider observables $V=V\lp(p_1,\ldots,p_n\rp)$, which are functions of the final-state momenta $p_1,\ldots,p_n$, and
that satisfy two basics properties:
\begin{enumerate} \label{enu:global_event_shape}
\item {\bf infrared and collinear safety}: an event shape is IRC safe if its value $v$ does not change in the presence of an arbitrary number of strictly soft or collinear emissions (see section~\ref{sec:ir_divergences}).
\item {\bf globalness}: an observable is defined to be global if it is sensitive to radiation anywhere in phase-space.
\end{enumerate}
As already discussed in section~\ref{sec:ir_divergences}, the first condition ensures cancellation of soft and collinear singularities between real and virtual corrections, so that the observable $v$ can be computed in (fixed-order) perturbation theory.
Moreover, it is also possible to show that non-perturbative corrections to event-shape distributions, such as the ones due to the hadronization process, are suppressed by inverse powers of the hard scale (see e.g.\ Refs.~\cite{Dokshitzer:1995zt,Dokshitzer:1995qm,Dokshitzer:1997ew, Dokshitzer:1997iz,Dokshitzer:1998pt,Dokshitzer:1998qp, Salam:2001bd,Dasgupta:2007wa}). 
Many observables of phenomenological interest, especially at hadron colliders do not obey the second condition and they are therefore termed \emph{non-global}~\cite{Dasgupta:2001sh}. Their all-order structure is more intricate and will be discussed in section~\ref{sec:non_global}.

To ensure the feasibility of resummation, the two conditions above have to be made more restrictive, as specified in Refs.~\cite{Banfi:2001bz,Banfi:2003je,Banfi:2004yd,Banfi:2004nk,Banfi:2010xy}. In fact, one needs to impose:
\begin{enumerate} \label{enu:global_event_shape_caesar}
\item {\bf \emph{recursive} infrared and collinear safety}: an event shape is rIRC safe if two conditions are met
\begin{enumerate}
\item in the presence of multiple soft and/or collinear emissions,
the observable should have the same scaling properties as with just one of them;
\item for sufficiently small values $V=v$ of the observable, there exists some $\e \ll 1$ such that $\e$ can be chosen independently of $v$ and emissions below $\e v$ do not significantly contribute to the observable.
\end{enumerate}
\item {\bf continuous globalness}: an observable is defined to be continuously global if its scaling with respect to the transverse momentum of a soft and/or collinear emission is the same everywhere in phase-space.
\end{enumerate}
We refer the interested Reader to Ref.~\cite{Banfi:2004nk} for a detailed discussion.

In our analysis, it will prove very convenient to consider the cumulative distribution for an event shape, 
defined as integral of the differential distribution up to the value $V=v$, normalized by the total cross-section:
\begin{eqnarray}\label{eq:thrust_sigma_def}
\Sigma(v)&\equiv&\frac{1}{\sigma_0}\int_0^v \d V \;  \frac{\d \sigma}{\d V}.
\end{eqnarray}
In the next section we consider an explicit example for such an event shape, namely thrust T~\cite{Farhi:1977sg}, defined in Eq.~(\ref{eq:thrust_def}). Many more event shape observables exist and we refer to~\cite{jones,Ford:2004dp,Banfi:2004nk,Banfi:2010xy} for an exhaustive list.

We adopt the convention that the logarithmic accuracy of a calculation is determined with respect to the logarithm of $\Sigma$, which after resummation can be written in exponentiated form as
\begin{equation}\label{eq:log_counting_def}
\Sigma(v)\propto\exp\left\{L\,g_{1}\lp(\as L\rp)+g_{2}\lp(\as L\rp)+\as g_{3}\lp(\as L\rp)+\ldots\right\}\,,
\end{equation}
where $L\equiv\ln\frac{1}{v}$. A leading logarithmic (LL) calculation then determines $g_{1}$, which resums contributions $\as^n L^{n+1}$ in $\ln \Sigma$ to all orders in perturbation theory. In this review we will be mostly considering next-to-leading logarithmic (NLL) accuracy, which corresponds to computing $g_1$ and $g_2$, i.e. resumming all $\as^n \ln^{n} \frac{1}{v}$ contributions to $\ln \Sigma$. Thus the large parameter which is resummed is $\as L\lesssim 1$. For completeness we remind that also other conventions can be adopted. In particular, for observables which do not exponentiate, it is possible to reorganize the perturbative expansion in terms of powers of $\as L^{2}$. The first term in this convention, also known as tower expansion~\cite{Catani:2000xk}, sums all double-logarithmic terms $(\as L^{2})^{n}$, the second one the terms $\as L(\as L^{2})^{n-1}$, and so on, assuming that the large parameter is $\as L^{2}\lesssim 1$.

\subsection{The thrust distribution}\label{sec:thrust}
In order to make this exposition as pedagogical as possible, while still discussing an observable of phenomenological relevance, we begin by considering as a concrete example the event shape thrust~\cite{Farhi:1977sg}. The NLL resummation of thrust, and other related event shapes, was first performed in Ref.~\cite{catani}, which we closely follow in this presentation.

\begin{figure}
\begin{center}
\includegraphics[scale=0.7]{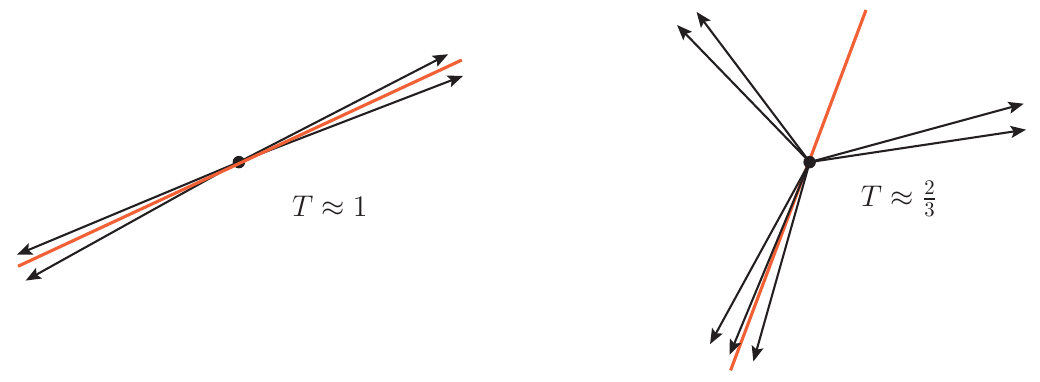}
\caption{Typical thrust values for back-to-back two-jet configurations (left) and three-jet final states (right).}\label{fig:thrust}
\end{center}
\end{figure}

Given a collection of final-state momenta $\{p_i\}=\{(E_i,\vec{p_i})\}$ we define
\begin{equation} \label{eq:thrust_def}
T= \max_{\vec{n}} \frac{\sum_i |\vec{p_i} \cdot \vec{n}|}{\sum_i |\vec{p_i}|}=
\max_{\vec{n}} \frac{\sum_i |\vec{p_i} \cdot \vec{n}|}{Q},
\end{equation}
where the second equality holds if all particles are massless, i.e.\ $p_i^2=0$ and $Q$ is the center-of-mass energy of the colliding electrons. The three-vector $\vec{n}=\vec{n}_T$ that maximizes the sum in Eq.~(\ref{eq:thrust_def}) is called the thrust axis. As depicted in Fig.~\ref{fig:thrust}, thrust measures how uniform radiation is distributed in the event, with $T \simeq 1$ indicating an event which is two-jet like. For convenience, the variable $\tau =1 -T$ is often introduced.

Before discussing the resummation of the thrust distribution, let us analyze its kinematics in the presence of soft and/or collinear radiation.  Let us call $P_{\vec{n}_T}$ the plane orthogonal to the thrust axis $\vec{n}_T$. This plane divides the event into two hemispheres $S_1$ and $S_2$. We also define hemisphere momenta:
\begin{eqnarray}
q_1&=& \sum_{i \in S_1}p_i=z_1 p+ q_{t1} +\bar z_1 \bar p\nonumber \\
q_2&=& \sum_{i \in S_2}p_i=z_2 p+ q_{t2} +\bar z_2 \bar p,
\end{eqnarray}
where $p,\bar p$ are lightlike momenta. We now state two properties of the thrust axis $\vec{n}_T$~\cite{catani}, namely that no $\vec{p_i}$ lies in $P_{\vec{n}_T}$ and that the hemisphere three-momenta are aligned with $\vec{n}_T$, i.e.\ $q_{t1}=q_{t2}=0$. Note that $q_i^2 \neq 0$. Thanks to these properties, and using energy-momentum conservation together with the above Sudakov parametrization we have
\begin{eqnarray}\label{eq:thrust_mass_kin}
\tau&=& 1- \frac{1}{Q} \left(|\vec{q}_1\cdot \vec{n}_T|+|\vec{q}_2\cdot \vec{n}_T| \right)\nonumber\\
&=&1- \sqrt{1-2 \left(\frac{q_1^2}{Q^2}+\frac{q_2^2}{Q^2}  \right)+
\left(\frac{q_1^2}{Q^2}-\frac{q_2^2}{Q^2}  \right)^2} = \frac{q_1^2}{Q^2}+\frac{q_2^2}{Q^2} + \ord \left(\frac{q_i^2 q_j^2}{Q^4} \right) \nonumber\\
&=& \sum_{i \in S_1} V_1(k_i)+\sum_{i \in S_2} V_2(k_i)+ \ord \left(\frac{q_i^2 q_j^2}{Q^4} \right)
=\sum_i V(k_i),
\end{eqnarray}
where in the last line all sums run over soft and/or collinear emissions, which we have labelled with momenta $k_i$, and $V_1(k_i)$, $V_2(k_i)$ are their contributions to the respective hemisphere mass.
Note that we have neglected recoil effects, which are beyond NLL accuracy~\cite{catani,Banfi:2004yd,Monni:2011gb}.
Thus, we have managed to turn the task of resumming the thrust distribution, into the resummation of the invariant masses of the two hemispheres.  Moreover, the second and third last lines of Eq.~(\ref{eq:thrust_mass_kin}) show that to the accuracy we are working at, the hemisphere mass and thrust are additive observables, i.e. the contribution of $n$ emission is the sum of the contributions of each emissions. 

We can now turn our attention to the actual resummation of the thrust distribution.
We start by considering the real-emission contribution to the cumulative distribution Eq.~(\ref{eq:thrust_sigma_def}), in the appropriate collinear limit. At NLL accuracy it can be written as a product of independent angular ordered emissions, as given by the generating functional derived in the previous section:
\begin{eqnarray}\label{eq:thrust_real}
W^R&=& \sum_{n=0}^\infty \frac{1}{n!} \prod_{i=1}^n \int \d z_i \frac{\d k_{ti}^2}{k_{ti}^2} \frac{\d \phi_i}{2 \pi} 2 C_F \frac{\as(k_{ti})}{2\pi} p_{gq}(z_i)
\Theta \left( \tau - \sum_{i=1}^n V(k_i)\right) \nonumber \\
\end{eqnarray}
where the transverse momenta $k_{ti}$ are measured with respect to the thrust axis.  We have introduced the splitting function $p_{gq}(z)=\frac{2-2z +z^2}{z}$ and the $n!$ accounts for ordering of the emitted gluons. The strong coupling $\as$ is evaluated in the so called CMW scheme~\cite{Catani:1990rr}
\begin{eqnarray}\label{eq:CMW}
\as= \as^{\MSb}\left(1+\frac{\as^{\MSb}}{2 \pi} K \right), \; {\rm with} \quad K= \ca \left( \frac{67}{18}-\frac{\pi^2}{6}\right)-\frac{5}{9}\nf,
\end{eqnarray}
i.e.\ it accounts inclusively for secondary branchings at NLL level. Note that $K$ is the coefficient of the two-loop cusp anomalous dimension\footnote{The dynamics of soft radiation can be computed considering correlators of Wilson lines with a cusp. Their renormalization leads to the cusp anomalous dimension, which controls the renormalization group evolution of these amplitudes~\cite{Polyakov:1980ca,Brandt:1981kf,Korchemsky:1992xv}. In the CMW scheme the first subleading contribution associated with soft emissions is accounted for with a redefinition of the strong coupling, allowing to capture the full NLL dependence.}. Factorization properties of Eq.~(\ref{eq:thrust_real}) appear to be spoiled by the presence of the $\Theta$-function that constrains the measured value of thrust and prevents us from summing the series into an exponential factor. As it is often the case in this kind of situation, the kinematic constraint can be diagonalized by considering a conjugate space. In particular, we introduce the following integral representation:
\begin{equation}\label{eq:transf_add}
\Theta \left( \tau - \sum_{i=1}^n V(k_i)\right)= \int \frac{\d \nu}{2 \pi i \nu} e^{\nu \tau} \prod_{i=1}^n e^{-\nu V(k_i)},
\end{equation}
where the dependence of $V$ on $p$ and $\bar p$ is understood.
Thus, by inserting the above expression into Eq.~(\ref{eq:thrust_real}), we are able to perform the sum over the number of emissions and we obtain
\begin{eqnarray}\label{eq:thrust_real_cntd}
W^R&=& \int \frac{\d \nu}{2 \pi i \nu} e^{\nu \tau}  \sum_{n=0}^\infty \frac{1}{n!} \prod_{i=1}^n \int \d z_i \frac{\d k_{ti}^2}{k_{ti}^2} \frac{\d \phi_i}{2 \pi} 2 C_F \frac{\as(k_{ti})}{2\pi} p_{gq}(z_i)  e^{-\nu V(k_i)} \nonumber \\
&=& \int \frac{\d \nu}{2 \pi i \nu} e^{\nu \tau}  \exp \left[ \int \d z \frac{\d k_{t}^2}{k_{t}^2} \frac{\d \phi}{2 \pi} 2 C_F \frac{\as(k_{t})}{2\pi} p_{gq}(z)  e^{-\nu V(k)} \right]
\end{eqnarray}
We now have to consider virtual corrections. In the soft or collinear limit, these are equal and opposite to the real-emission one. However, no observable constraint is present because the kinematics is fixed to Born level, leading straightforwardly to exponentiation. Thus, putting together real and virtual corrections, the all-order cumulative distribution reads
\begin{eqnarray}\label{eq:thrust_start}
\Sigma(\tau)&=& \int \frac{\d \nu}{2 \pi i \nu} e^{\nu \tau} \exp \left[ \int \d z \frac{\d k_{t}^2}{k_{t}^2} \frac{\d \phi}{2 \pi} 2 C_F \frac{\as(k_{t})}{2\pi} p_{gq}(z)  \left( e^{-\nu V(k)}-1\right) \right]. \nonumber\\
\end{eqnarray}
We now want to compute the resummed exponent to a fixed logarithmic accuracy. In this example, and for most of this review, we are working at NLL accuracy. To this purpose, we evaluate the integrals over the gluon phase-space, making the following approximations:
\begin{enumerate}
\item we evaluate the running coupling with the two-loop QCD $\beta$ function;
\item we also note that the factor $\left( e^{-\nu V(k)}-1\right)$ essentially acts as a restriction on the allowed phase space. To NLL accuracy we can replace it with~\cite{Catani:1989ne} $$\left( e^{-\nu V(k)}-1\right) \to -\Theta\left( V(k)-e^{-\gamma_E}\nu^{-1}\right)\,,$$ where $\gamma_E$ is the Euler constant. The generalization of this procedure to higher logarithmic accuracy is discussed in Appendix~A of Ref.~\cite{Catani:2003zt};
\item we integrate the finite part of the splitting function down to $z=0$, i.e.\ $$B_q=\int_0^1 \d z \left(p_{gq}(z)-\frac{2}{z} \right)=-\frac{3}{2}\,.$$
\end{enumerate}
We obtain
\begin{eqnarray} \label{eq:resum_nu}
\Sigma(\tau)&=& \int \frac{\d \nu}{2 \pi i \nu} e^{\nu \tau} \exp \left[\frac{1}{\as} g_1(\lambda_\nu) +g_2(\lambda_\nu) \right],
\end{eqnarray}
with $\lambda_\nu=\as \beta_0 \ln \nu$ and $\as=\as(Q)$. The function $g_1$ resums LL contributions, while $g_2$ NLL ones:
\begin{eqnarray} \label{eq:g1_g2}
g_1(\lambda)&=&- \frac{\cf}{ \pi \beta_0^2} \left[(1- 2\lambda_\nu) \ln (1- 2\lambda_\nu) - 2 (1- \lambda_\nu) \ln (1- \lambda_\nu) \right], \nonumber\\
g_2(\lambda)&=& -\frac{\cf K}{2 \pi^2 \beta_0^2}\left[ 2 \ln (1- \lambda_\nu) -  \ln (1- 2\lambda_\nu) \right]+\frac{\cf B_q}{\pi \beta_0} \ln(1-\lambda_\nu)
\nonumber\\
&+&\frac{\cf \beta_1}{\pi \beta_0^3}\Big[ 2 \ln (1- \lambda_\nu) -  \ln (1- 2\lambda_\nu) + \ln^2 (1- \lambda_\nu)
\nonumber\\ &-&  \frac{1}{2}  \ln^2 (1- 2\lambda_\nu) \Big]
- 2\frac{\cf \gamma_E}{\pi \beta_0}\left[ \ln (1- \lambda_\nu) -  \ln (1- 2\lambda_\nu) \right].
\end{eqnarray}
In order to perform the inverse Mellin transform, we expand the exponent in powers of $\ln \nu$ around $\ln \nu=-\ln \tau$
\begin{eqnarray} \label{eq:resum_tau}
\Sigma(\tau)&=& \exp \left[\frac{1}{\as} g_1(\lambda) +g_2(\lambda) \right] \int \frac{\d \nu}{2 \pi i \nu} e^{\nu \tau+\frac{1}{\as}
\left(\partial_L g_1(\lambda)\right) \left( \ln \nu +\ln \tau \right)  +\dots }, \nonumber \\
&=&\exp \left[\frac{1}{\as} g_1(\lambda) +g_2(\lambda) \right] \Bigg / \Gamma \left(1-\as^{-1}\partial_L g_1(\lambda)\right).
\end{eqnarray}
with $\lambda=\as \beta_0 L$ and $L= -\ln \tau$ and the dots stay for higher derivatives, which give rise to sub-leading contributions. Thus, we have reached the main result of this section: Eq.~(\ref{eq:resum_tau}) represents the NLL resummed cumulative distribution for thrust.
The result can be upgraded to sometimes called NLL$^\prime$ accuracy by considering the $\ord(\as)$ constant contributions, schematically $\Sigma\to (1+\as g_{0,1}) \Sigma$.

Some comments about this result are in order. Eq.~(\ref{eq:resum_tau}) is expected to capture the dominant physical effects in the region where the logarithms are large, i.e.\ $\as L^2\sim1$, but the use of perturbative QCD is still justified. Note that the functions $g_i$ have logarithmic branch cuts starting from $\lambda =1/2$, which corresponds to $\tau=e^{-\frac{1}{2 \as \beta_0}}$. This singularity originates from the Landau pole of the QCD running coupling and heralds the breakdown of the perturbative approach. 
%
Even before reaching this singularity, as the value of $\tau$ decreases, it becomes sensitive to emissions with transverse momentum in the non-perturbative region, and non-perturbative power corrections must be taken into account. 
A thorough discussion of these effects goes beyond the purpose of this review and we refer the interested reader to the original literature for more details~\cite{Manohar:1994kq,Korchemsky:1994is,Dokshitzer:1995qm,Korchemsky:1998ev,Gardi:1999dq,Hoang:2007vb}. Let us just briefly mention that several approaches exist. Their basic common idea is to refrain from doing the substitution (ii) described above in Eq.~(\ref{eq:thrust_start}), and instead to split the integral in the exponential into a perturbative and a non-perturbative part. In the so-called \textit{dispersive model}~\cite{Dokshitzer:1995zt,Dokshitzer:1995qm,Dokshitzer:1997ew} the non-perturbative part is treated by defining an effective coupling, which is supposed to be finite in the infrared region and which can be written in terms of a dispersive relation. Recently this model was extended to match NNLL+NNLO accuracy for thrust~\cite{Gehrmann:2012sc}. Other alternatives are the \textit{single dressed gluon approximation}~\cite{Gardi:1999dq,Gardi:2000yh,Gardi:2001ny}, which assumes the existence of a reordering of the perturbative series in a so-called skeleton expansion, or the introduction of a \textit{shape function}~\cite{Korchemsky:1998ev,Korchemsky:1999kt,Korchemsky:2000kp}, which is a non-perturbative function which admits an operator definition in terms Wilson lines, and was recently extended and used in the context of SCET~\cite{Hoang:2007vb,Abbate:2010xh}. The inclusion of non-perturbative effects is important in the context of strong coupling determinations, where also their universality and the effect of finite mass corrections can be tested~\cite{Akhoury:1995sp,Dokshitzer:1997iz,Salam:2001bd,Lee:2006nr,Mateu:2012nk}.

\begin{figure}
\begin{center}
\includegraphics[scale=0.8]{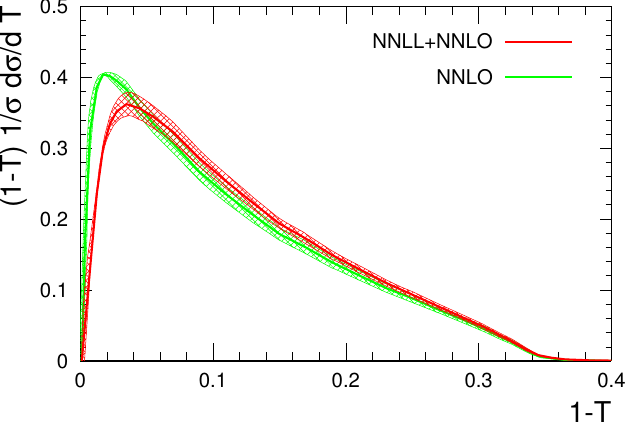}
\caption{The distribution of the thrust event shape at NNLL matched to NNLO, in red, compared to the fixed-order (NNLO) result, in green. The bands indicate the theoretical uncertainty. Plot taken from Ref.~\cite{Monni:2011gb}.}\label{fig:thrust_res}
\end{center}
\end{figure}

The resummed result on its own is also inadequate for describing the region where $\tau$ is an order one quantity, because the soft and/or collinear approximation on which resummed calculations are based upon breaks down. This is the domain of validity of fixed-order calculations, which can be matched to the resummation in order to obtain solid theoretical predictions over a vast range of $\tau$. As an example, in Fig.~\ref{fig:thrust_res}, we show the matched distribution for thrust. The calculation is actually performed at a higher accuracy than the one presented above: the resummation is performed to NNLL and it is matched to a NNLO fixed-order calculation~\cite{Monni:2011gb}. A similar result with part of the NNNLL contribution has been computed also in SCET~\cite{Becher:2008cf}.

It is also useful to discuss the limitations of the framework in which the above calculations has been carried out, as well as possible extensions. Firstly, the original derivation of the all-order thrust distribution was based on the coherent branching algorithm described in section~\ref{sec:resummation}, which exploits color-coherence of azimuthally-integrated matrix elements that are relevant for one- and two-jet observables~\cite{Catani:1990rr,catani}.
A framework that enables one to perform resummed calculations in the presence of many hard legs is desirable but not straightforward beyond LL accuracy. While single-logarithmic terms arising from hard collinear emissions are still captured by a probabilistic method, the correct description of large-angle soft radiation requires to go beyond it.
 This issue was addressed by different groups. For instance, NLL resummation of near-to-planar 3-jet events was performed in Ref.~\cite{Banfi:2000si}, while  general frameworks to resum soft large-angle logarithms in processes with many legs was worked out in Refs.~\cite{Contopanagos:1996nh,Kidonakis:1997gm,Kidonakis:1998bk,Kidonakis:1998nf} and Ref~\cite{Bonciani:2003nt}. The final result is actually remarkably simple and it amounts to consider soft emissions from each of the dipole that constitute the system of $n$ hard partons, weighted by an operator describing the color exchange. We will come back to this point in the next section.

The second limitation that we wish to stress has to do with the nature of the observable that we wish to resum. Central to the above analysis was the identification of the integral transform to diagonalize the observable Eq.~(\ref{eq:transf_add}), which relied upon the additive nature of thrust. Different, and possibly more complicated observables, require multiple integral transforms, restricting the set of resummable observables to the ones for which a theorist's ingenuity has not failed to provide the correct integral transform. This is not a pleasant situation and a more automated method is highly desirable.
 Such a framework was devised in Refs~\cite{Banfi:2001bz,Banfi:2010xy,Banfi:2004yd,Banfi:2003je,Banfi:2004nk} and will be the subject of the next section.

\subsection{The \caesar approach}\label{sec:caesar}
\caesar~\cite{Banfi:2010xy,Banfi:2001bz,Banfi:2004yd,Banfi:2003je,Banfi:2004nk} is a framework (and a computer program) that allows one to perform the resummation of a large class of observables, namely global event shapes, to NLL accuracy.
We begin this discussion by considering, as in the previous section, processes which at Born level feature two hard massless partons (plus eventually color singlets, e.g.\ photons, Higgs or electroweak bosons) and we denote the set of Born momenta with $\{p\}$~\footnote{In principle  one should consider the set of momenta $\{\tilde{p}\}$ after recoil, but this effect is beyond NLL.}. We consider positive-definite observables $V$ that measure the difference in the energy-momentum flow of an event with respect to the Born configuration. In particular, we focus on observables that vanish when evaluated on Born configurations, $V\left( {\{p}\}\right)=0$. The previously discussed variable $\tau=1-T$, which vanishes in the 2-jet limit, is an example of such observables.

We consider the cumulative distribution for the event shape $v$ in the presence of $n$ soft or collinear emissions $k_i$
\begin{eqnarray}\label{eq:sigma_caesar_start}
\Sigma(v)&=&e^{- \int [\d k] M^2(k)} \nonumber \\ &\times &\sum_{n=0}^{\infty} \frac{1}{n!}
  \int\prod_i [\d k_i] M^2(k_i)\,\Theta\left(v-V(\{k_i\})\right).
\end{eqnarray}
Henceforth, the dependence of $V$ upon the Born momenta is understood, i.e.\ $V(\{k_i\})\equiv V(\{p\},\{k_i\})$ The first term corresponds to virtual corrections to the Born process, which exponentiate in the soft or collinear limit, as previously discussed.
$M^2$ is the matrix element squared for the real emission of $n$ soft or collinear partons off the two hard partons.
The key idea is to further divide the real-emission phase-space by introducing a resolution parameter $\e$. Resolved emissions give $V>\e v$, while unresolved ones $V<\e v$. Because of rIRC safety, unresolved emissions do not significantly contribute to the measured value of the event shape and therefore, like virtual corrections, they exponentiate
  \begin{eqnarray}\label{eq:sigma_caesar_cntd}
\Sigma(v)&=&e^{- \int [\d k] M^2(k) \big[1- \Theta\left(\e v-V(k)\right)\big]}
\sum_{n=0}^{\infty} \frac{1}{n!}
  \int_{\e v}^v \prod_i [\d k_i] M^2(k_i), 
  \end{eqnarray}
where we have introduced the shorthand notation
\begin{eqnarray}
  \int_{\e v}^v \prod_i [\d k_i] M^2(k_i)&=&   \int \prod_i [\d k_i] M^2(k_i)\Theta\left(v-V(\{k_i\})\right)
  \nonumber \\ && \times \Theta\left(V(\{k_i\})-\e v\right)
\end{eqnarray}
It is now natural to combine virtual and unresolved contributions:
\begin{eqnarray}\label{eq:sigma_caesar_final}
\Sigma(v)&=&e^{ - \int [\d k] M^2(k) \Theta\left(V(k)-v \right)}  \mathcal{F}(v),
\end{eqnarray}
where we have introduced
\begin{eqnarray}  \label{eq:F}
  \mathcal{F}(v)&=&
  e^{-\int_{\e v}^{v}[\d k] M^2(k)}\sum_{n=0}^{\infty}\frac{1}{n!}\int_{\e v}^v \prod_{i=1}^n [\d k_i] M^2(k_i).
\end{eqnarray}
The first contribution only depends on one emission and, once evaluated to a fixed-logarithmic accuracy, it is the generalization of the resummed exponent in the resummation of thrust Eq.~(\ref{eq:resum_tau}). It has the following physical interpretation: given an emission that sets the value $v$ of the event shape, it vetoes further emissions which would contribute to the event shape more than $v$. The second term instead takes into account the effect of multiple emissions that equally contribute to the event shape.

We now briefly describe how to evaluate both the resummed exponent and the function $\mathcal{F}$ to NLL.
Consider a  single emission with momentum $k$, which is soft and collinear to a hard leg $l$.
We can parametrize the value of the event shape in the presence of this emission as follows
\begin{equation}\label{caesar-obs}
V\left(k\right)= d_l \left(\frac{k_t^{(l)}}{Q} \right)^{a}e^{-b_l \eta^{(l)}}g_l\left(\phi^{(l)} \right),
\end{equation}
where $k_t^{(l)}$, $\eta^{(l)}$ and $\phi^{(l)}$ denote transverse momentum, rapidity and azimuth of the emission, all measured with respect to parton $l$. $Q$ is the hard scale of the process which can be set, for instance, equal to the partonic centre of mass energy, i.e.\ $Q=\sqrt{s}$. Note that the requirement of continuous globalness requires all the $a$ coefficients to be the same, while IRC safety imposes $a>0$.

\begin{figure}
\begin{center}
\includegraphics[scale=0.7]{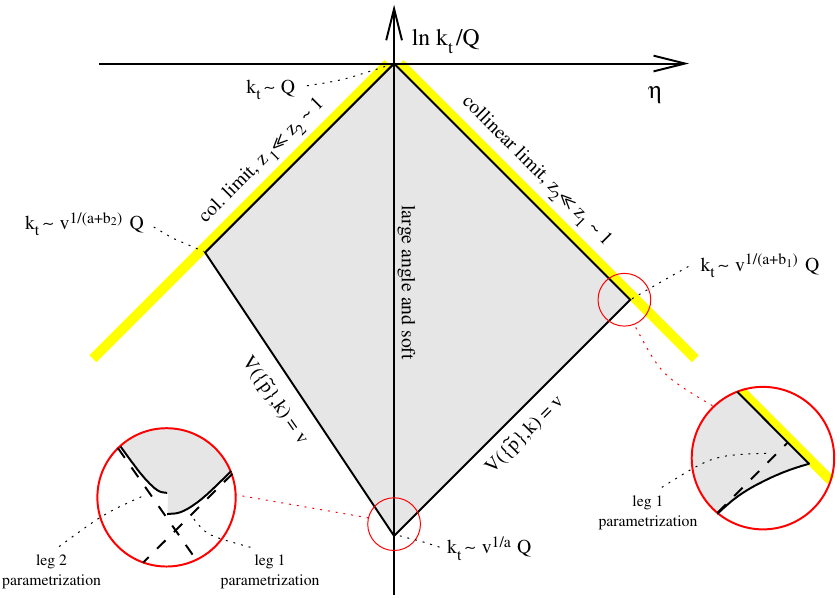}
\caption{The emission phase-space in the ($\eta, \ln k_t/Q$) plane, as parametrized by the \caesar formula Eq.~(\ref{caesar-obs}.)
We note that, in the soft and collinear limit, emissions are uniformly distributed in this plane.
Figure taken from Ref.~\cite{Banfi:2004yd}.}\label{fig:phase-space}
\end{center}
\end{figure}

Thus far we have considered, mainly for simplicity, only the case of an underlying Born with two hard partons. However, as we are about to evaluate the resummed exponent, we can lift this restriction and consider an ensemble of $m$ hard partons.
The main complication we have to deal with is the presence of many color configurations, which renders the virtual correction matrices in color space. Therefore the exponentiation in Eq.~(\ref{eq:sigma_caesar_final}) has to be understood in a formal way. The calculation of $\mathcal{F}$ instead does not change at this logarithmic accuracy. We have~\footnote{Beyond NLL accuracy considered here, one needs to introduce path-order exponentials in order to deal with the exponentiation of non-commuting matrices.}
\begin{eqnarray} \label{eq:sigma_color}
\Sigma(v) =\frac{1}{\sigma_0}\langle M_0 | e^{- \bf R^\dagger} e^{- \bf R}|M_0 \rangle \mathcal{F},
\end{eqnarray}
where $| M_0\rangle$ is the Born amplitude, i.e.\ $\sigma_0=\langle M_0 |M_0\rangle$. The real part of the resummed exponent has a structure similar to the one we have encountered for thrust
\begin{eqnarray}\label{eq:rad_matrix}
{\rm Re}\,{\bf R}&=& \sum_{{\rm dipoles}\, ij} \sum_{{\rm legs }\,  l \in ij} (- \TiTj)  \int \frac{\d z}{z} \frac{\d k_{t}^2}{k_{t}^2} \frac{\d \phi_i}{2 \pi} \frac{\as(k_t)}{2 \pi} z^{(l)} p \left(z^{(l)} \right)\Theta(\eta)
\nonumber \\ &\times&
\Theta \left( d_l \left(\frac{k_t^{(l)}}{Q} \right)^{a}e^{-b_l \eta^{(l)}}g_l\left(\phi^{(l)}\right)-v \right),
\end{eqnarray}
where $\eta=\frac{1}{2} \ln \frac{z^{(i)}}{z^{(z)}}$ and $p$ is the appropriate reduced splitting function. The integral above corresponds to the shaded area in Fig.~\ref{fig:phase-space}. Note that we have arbitrarily separated the integration at $\eta=0$, identifying a region where emissions are collinear to leg $i$ and leg $j$ respectively. The precise position of this boundary is beyond NLL accuracy.
Moreover, we have also made use of the color product $\TiTj$ operators introduced in section~\ref{sec:gen_func_QCD} and applied in the context of resummation for the first time in Ref.~\cite{Bonciani:2003nt}.

The resummed exponent ${\bf R}$ has also an imaginary part known as Coulomb phase (see discussion after Eq.~(\ref{eq:loopintsoft})) which is proportional to $\TiTj$, where $i,j$ are both final-state partons (or both initial-state partons in the case of hadron collision). This phase cancels in the product of exponentials when $m<4$, while gives a physical contribution to the distributions in processes with four or more hard legs.

It is now relatively straightforward to compute the one-emission integrals; using color-conservation Eq.~(\ref{eq:color_conservation}) we are able to cast the $\caesar$ resummed formula Eq.~(\ref{eq:sigma_caesar_final}) in the same form as Eq.~(\ref{eq:resum_tau}). The LL function $g_1$ that appears in the resummed exponent generalizes to
\begin{eqnarray}\label{caesar-g1}
g_1(\as L)&\equiv&\sum_{l=1}^m g_{1,l}(\as L)=
-\sum_{l=1}^m  \frac{C_l}{2 \pi \beta_0^2 b_l}\Big[
(a-2 \lambda) \ln \left( 1-\frac{2\lambda}{a}\right)\nonumber \\ &&-(a+b_l-2 \lambda)\ln \left(1-\frac{2 \lambda}{a+b_l}\right)\Big],
\end{eqnarray}
The above LL result consists of a sum over all hard partons, the dependence on the color is trivial and only enters through the Casimir of each leg $l$, ($C_F$ for a quark leg, $C_A$ for a gluon leg). The thrust result is recovered by considering only one $q\bar q $ dipole and by setting $a=b_l=d_l g_l =1$, $l=1,2$.
The result for the NLL function $g_2$ has a richer structure:
\begin{eqnarray}\label{g2-caesar}
g_2(\as L)&= &-\sum_{l=1}^m C_l \left[ \frac{r^{(2)}_l}{b_l}+ B_l \, T \left( \frac{L}{a+b_l} \right) \right]+\partial_L \left[ \as^{-1}g_{1,l}(\as L)\right]D_l
\nonumber\\
&+& \ln \mathcal{S}\left(T\left(L/a\right)\right) +\ln \mathcal{F}.
\end{eqnarray}
The first term in the square brackets in Eq.~(\ref{g2-caesar}) contains the two-loop contribution to the DGLAP splitting function in the soft limit and to the QCD $\beta$-function
\begin{eqnarray}\label{caesar-g2-r2}
r_l^{(2)}&=& \frac{K}{4 \pi^2 \beta_0^2}\left[ \left( a+b_l\right) \ln \left(1-\frac{2 \lambda}{a+b_l}\right)-a \ln\left(1-\frac{2 \lambda}{a}\right) \right] \nonumber \\
&+&\frac{\beta_1}{2 \pi \beta_0^3}\left[ \frac{a}{2} \ln^2 \left(1-\frac{2\lambda}{a}\right)-\frac{a+b_l}{2}\ln^2\left(1-\frac{2 \lambda}{a+b_l}\right) \right.
\nonumber \\ &+& \left. a \ln\left(1-\frac{2 \lambda}{a}\right)-(a+b_l)\ln\left(1-\frac{2 \lambda}{a+b_l}\right)\right].
\end{eqnarray}
The second term in the square brackets in Eq.~(\ref{g2-caesar}) instead captures hard collinear emissions to a quark leg ($B_q=-\frac{3}{4}$) or to a gluon leg ($B_g=-\pi \beta_0$); we have also introduced
\begin{eqnarray}\label{Tfunction}
T(L) &=& \frac{1}{\pi \beta_0} \ln\frac{1}{1-2 \as \beta_0 L}, \\
D_l& =& \ln d_l\left( \frac{Q}{2 E_l}\right)^{b_l}+\int_0^{2\pi }\frac{\d \phi}{2 \pi} g_l(\phi),
\end{eqnarray}
where $E_l$ is the energy of leg $l$.
The contribution $\mathcal{S}$ instead captures the effect of large-angle soft emission. At NLL, it is the only contribution for which we have to keep track of the matrix structure in color space described above. In order to explicitly evaluate this contribution, one needs to fix a color basis, compute the representation of the color products $\TiTj$ and the color-decomposed Born amplitude. Although straightforward, these steps can be tedious and laborious because the dimensionality of the basis quickly increases with the number of hard legs. For this reason, until very recently, only calculations with up to four hard legs were performed~\footnote{Refs.~\cite{Forshaw:2006fk, Forshaw:2008cq, Forshaw:2009fz,DuranDelgado:2011tp} considered the resummation of $2\to3$ scattering processes, in the limit where one of the final-state partons was a soft gluon.}.

We now turn our attention to  $\mathcal{F}$, which describes the effect of multiple emissions. This contribution only starts at single-logarithmic level, therefore, in order to perform a NLL calculation, we can consider all the emissions in Eq.~(\ref{eq:F}) to be soft and collinear and widely separated in rapidity. Moreover, the parametrization of the observable simplifies as well, because we can ignore the overall normalization $d_l g_l(\phi)$. These considerations lead to a simpler expression in the case of event-shape variables considered here (see Ref.~\cite{Banfi:2004yd} and Ref.~\cite{Banfi:2014sua} for details):
\begin{eqnarray}\label{eq:F_nll}
\mathcal{F}&=& e^{ \ln \frac{1}{\e}\left[ \as^{-1}g_{1,l}(\as L)\right]}
   \sum_{n=0}^{\infty}\frac{1}{n!} \prod_{i=1}^n
    \int_{\e}^{\infty} \frac{\d \zeta_i}{\zeta_i}\int_0^{2\pi}
   \frac{\d \phi_i}{2\pi} \sum_{l} \left[- \as^{-1}g_{1,l}(\as L)\right] \nonumber \\ &&\times
   \Theta\left(1-\lim_{v\to 0}\frac{V(\{ p\},\{k_i\})}{v}\right),
\end{eqnarray}
where $\zeta_i=v_i/v$ is the fractional contribution to the observable due to emission $i$. The above expression contains only NLL terms and it is suitable for numerical evaluation via a Monte Carlo generator.

Thanks to the \caesar framework, it has been possible to perform resummed calculations for a variety of event shapes, for which no explicit integral transform to diagonalize the observable was known. Moreover, as we shall see in the next section, this framework can be extended, with minor modifications, to event shapes in hadron collisions.

Finally, we note that in some cases further singularities can appear inside the physical region of the phase space. These singularities originate for example when the phase-space boundary for a given number of partons lies inside the one for a larger number of partons. In this case, if the observable is discontinuous in that point, logarithms will appear which have to be resummed to all orders to make the observable finite there. One speaks in this case of a Sudakov shoulder. These logarithms can also be resummed~\cite{Catani:1997xc}, and in the very general case of cuts applied in experimental analyses, this task is usually fulfilled by parton-shower Monte Carlo generators.

\subsection{Extension to lepton-hadron and hadron-hadron collisions}\label{eq:event shapes_hadron}
\begin{figure}
\begin{center}
\includegraphics[scale=0.7]{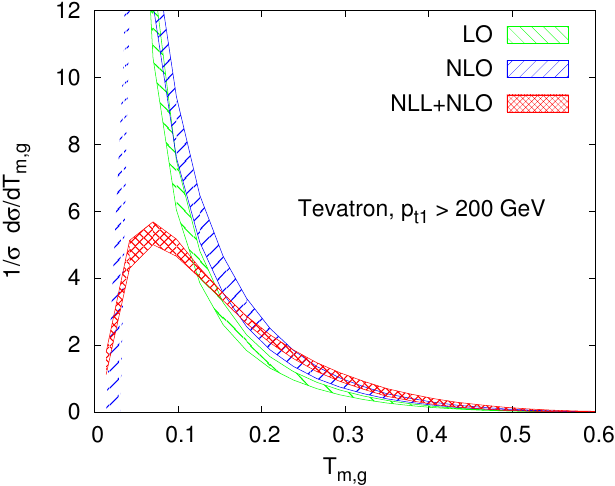}
\caption{
Distribution of the directly global thrust minor in $p \bar p$ collisions computed in three different approximations.
The fixed-order results are computed at LO (in green), and NLO (in blue). The resummed and matched distribution is instead computed with \caesar at NLL+NLO (in red). Plot taken from Ref.~\cite{Banfi:2010xy}.}\label{fig:caesar-plots}
\end{center}
\end{figure}
The above analysis for event shapes in lepton colliders can be extended to the case of lepton-hadron and hadron-hadron collisions (even though the framework we are discussing may break down at sufficiently high orders, see section~\ref{sec:factorization}). When computing cross-sections with one or two hadrons in the initial state, parton distribution functions (PDFs) are introduced in order to describe the long-distance non-perturbative physics of the initial state. PDFs obey the Dokshitzer-Gribov-Lipatov-Altarelli-Parisi (DGLAP) evolution equation~\cite{Gribov:1972ri,Dokshitzer:1977sg,Altarelli:1977zs}, which resums single logarithms due to the emissions of collinear partons off the incoming legs up to a factorization scale, which is usually taken to be of the order of the hard scale of the process $Q$.

However, the phase-space restriction that is imposed by measuring an event shape prevents us from inclusively integrating collinear emissions above $k_t \sim Qv^\frac{1}{a+b_l}$, because such emissions would result in too high a value for the event shape. Therefore, we have to correct the PDF scale choice in the Born cross-section (which only knows about $Q$) in order to account for this effect. This leads to a new contribution to the NLL function $g_2$:
\begin{equation}\label{eq:g2PDF}
\delta g_2= \sum_{l=1}^{n_{\rm initial}} \ln \frac{q^{(l)}(x_l, Q^2e^{-\frac{2 L}{a+b_l}})}{q^{(l)}(x_l, Q^2)},
\end{equation}
which effectively replaces the PDF $q^{(l)}(x,Q^2)$ evaluated at the hard scale, with the one evaluated at a lower scale, set by the event shape. We further note that the virtual part of collinear corrections are already accounted for by the $B_l$ term in Eq.~(\ref{g2-caesar}).
The actual proof of this result is not entirely straightforward and we refer the interested reader to Appendix~E of Ref.~\cite{Banfi:2004yd}.

Phenomenological studies of event shapes at lepton-hadron colliders using resummed perturbation theory have been reviewed in Ref.~\cite{Dasgupta:2003iq} (see Ref.~\cite{Kang:2013nha} for recent developments), while the hadron-hadron case was studied in Ref.~\cite{Banfi:2010xy}.
An example of resummed and matched result is shown in Fig.~\ref{fig:caesar-plots}, where the all-order (NLL+NLO) prediction for an event shape is compared to its LO and NLO approximations. The observable chosen as an example is the directly global thrust minor as measured in $p \bar p$ collisions.

A very interesting topic, both theoretically and phenomenologically, which would deserve a review on its own, is the resummation of the transverse momentum ($Q_T$) distribution of electro-weak final states, such as the Higgs boson or a lepton pair produced via the Drell-Yan mechanism.

The literature on $Q_T$ resummation is vast and since the seminal paper Ref.~\cite{Collins:1984kg}, there has been a continuous effort in producing accurate theoretical predictions that can describe the experimental data. For example, high logarithmic accuracy~\cite{Bozzi:2003jy,Bozzi:2005wk,Bozzi:2010xn,Catani:2010pd,Becher:2010tm,Catani:2013tia,Gehrmann:2014yya} has been achieved and computer programs that allow one to obtain NNLL+NLO predictions for the $Q_T$ distribution in case of colorless final states in hadron collision exist, e.g.~\cite{Ladinsky:1993zn,Landry:2002ix,Bozzi:2005wk,Bozzi:2010xn,Banfi:2012du,deFlorian:2011xf,deFlorian:2012mx}. Very recently first $Q_T$ resummation results were computed also for colored final states, and in particular for heavy quarks~\cite{Zhu:2012ts,Li:2013mia,Catani:2014qha}.

Novel observables such as the $a_T$~\cite{Vesterinen:2008hx} and $\phi^*$~\cite{Banfi:2010cf} variables that exploit angular correlations to probe similar physics as $Q_T$, while being measured with much better experimental uncertainty, have been introduced. This triggered theoretical studies to extend the formalism of $Q_T$ resummation to these observables~\cite{Banfi:2009dy,Banfi:2011dm,Banfi:2011dx,Banfi:2012du,Guzzi:2013aja} (for a brief review see Ref.~\cite{Marzani:2012hj}). The experimental resolution of $\phi^*$  is so good~\cite{Abazov:2010mk,Aaij:2012mda,Aad:2012wfa,Aad:2014xaa,Abazov:2014mza} that the theoretical uncertainty of the state-of-the-art NNLL+NLO calculation is much larger than the experimental one, calling for improved theoretical predictions.

An important issue in the context of Higgs physics is the ability of separating events according to their jet multiplicity. In particular, events where the Higgs is produced in association with $n$-jets are usually identified by vetoing additional radiation above a given threshold. Jet vetoes are usually applied to transverse momentum variables and therefore the all-order treatment of the 0-jet bin of Higgs cross-section is naturally related to the Higgs $Q_T$ spectrum itself~\cite{Banfi:2012yh}. This is actually true at low-logarithmic accuracy, while other effects, such as the ones due to the parton recombination to form a jet, become relevant at higher orders~\cite{Banfi:2012jm,Stewart:2013faa,Becher:2013xia}.

Thus far we have discussed the importance of resummation for exclusive processes. We also want to mention the fact that there are situations where resummation can become relevant even for inclusive cross-sections. Let us consider the ration between the typical hard scale of a process over the collider energy, $x=Q^2/s$. If $Q^2\sim s$, and consequently $x\to 1$, the final state is produced near threshold and logarithms of $1-x$ become large and may require resummation. As for the aforementioned $Q_T$ resummation, there exists an extensive literature on threshold resummation, see e.g.~\cite{Catani:1989ne,Sterman:1986aj,Forte:2002ni}. Here, we simply  report the state of the art, which is N$^3$LL for $2\to 1$ processes, such as deep inelastic scattering, Higgs production and Drell-Yan~\cite{Moch:2005ba,Moch:2005ky,Laenen:2005uz,Bonvini:2014joa,Catani:2014uta,Ahmed:2014cla,Ahmed:2014uya}. NNLL threshold resummation is also included in state-of-the-art theoretical predictions for the $t \bar t $ cross-section~\cite{Czakon:2009zw,Czakon:2013goa}.

In the opposite, high energy, limit of $x\to0$, small-$x$ logarithms become dominant. Such contributions, which are governed by the Balitsky-Fadin-Kuraev-Lipatov (BFKL) equation~\cite{Lipatov:1976zz,Fadin:1975cb,Kuraev:1976ge,Kuraev:1977fs,Balitsky:1978ic,Fadin:1998py}, contaminates both the evolution of the parton densities and of the partonic coefficient functions. The resummation of small-$x$ contributions to PDF evolution was investigated in the 1990s by more than one group to NLL, see, for example, Refs.~\cite{Salam:1998tj,Ciafaloni:1999yw,Ciafaloni:2003rd,Ciafaloni:2007gf} and Refs.~\cite{Ball:1995vc,Ball:1997vf,Altarelli:2001ji,Altarelli:2003hk,Altarelli:2005ni,Altarelli:2008aj}. The resummation of partonic coefficient functions is based on the so-called $k_t$-factorization theorem~\cite{Catani:1990xk,Catani:1990eg,Collins:1991ty,Catani:1993ww,Catani:1993rn,Catani:1994sq,Ball:2007ra,Caola:2010kv} and it is known to LL for an increasing number of cross-sections and distributions~\cite{Ball:2001pq,Hautmann:2002tu,Marzani:2008az,Harlander:2009my,Marzani:2008uh,Diana:2010ef} . Monte Carlo programs that incorporate small-$x$ contributions also exist, e.g.\ \cascade~\cite{Jung:2001hx} and \hej~\cite{Andersen:2011hs}.

We conclude this brief discussion by mentioning a case where the inclusion of all-order effects has proved itself crucial in the context of interpreting a discrepancy between theory and experimental data. Both the ATLAS and CMS collaborations have reported an excess in the inclusive $WW$ cross-section with respect to the Standard Model predictions~\cite{ATLAS:2012mec,atlasWWconf,Chatrchyan:2011tz,Chatrchyan:2013oev}. This has clearly sparked a lot of interests as a possible manifestation of new physics. However, careful analyses have shown that the excess can be largely explained within the Standard Model, if all-order resummation is taken into account when computing the measured fiducial cross-section~\cite{Meade:2014fca,Jaiswal:2014yba,Monni:2014zra,Wang:2015mvz}.

\subsection{Recent developments}\label{sec:event_shape_recent}
Before finishing this chapter, we would like to mention two recent developments in the resummation of event shapes.
As we discussed in detail,  the \caesar approach provides a highly-automated framework to compute global event shapes at NLL.
Moreover, the structure of soft singularities in multi-loop amplitudes has been extensively studied in the literature, e.g.~\cite{Catani:1998bh,Sterman:2002qn,Aybat:2006wq,Aybat:2006mz} and their all-order behavior turns out to be highly-constraint by symmetries~\cite{Gardi:2009qi,Becher:2009cu,Becher:2009qa}.

However, as we mentioned earlier, in order to correctly capture the logarithmic terms originating from emissions of soft gluon at large angles beyond LL, color bases have to be identified in order to explicitly compute color operators and Born amplitudes. Because of the increasing complexity of these bases, until recent, calculations were performed by-hand for relatively low multiplicities, e.g.\  $2\to 2$ and $2\to 3$ QCD scattering.
However, color-flow information is built in modern matrix element generators, such as \comix~\cite{Gleisberg:2008fv} or \madgraph~\cite{Alwall:2011uj}.
Recent work has solved this problem by constructing and implementing a framework that exploits this observation and allows for a calculation of large-angle soft logarithms in an highly automated way~\cite{Gerwick:2014gya}.
The algorithm constructs an appropriate color basis for the partonic process at hand, and evaluates color operators and the decomposition of Born amplitudes in that basis; color-ordered partial amplitudes are then evaluated using  \comix.
By merging this recent development together with the \caesar approach, NLL resummation in the presence of many hard jets, e.g. $2 \to 5$ QCD scattering, is now automated.

The second important recent development is the extension of the resummation to NNLL accuracy, which was achieved for global 2-jet event shapes in Ref.~\cite{Banfi:2014sua}. The resummed distribution can be written as
\begin{eqnarray}\label{eq:sigma_nnll}
\Sigma(v)= e^{\frac{1}{\as} g_1(\lambda)+g_2(\lambda)+\as g_3(\lambda)} \left[ \mathcal{F}(\lambda)+\delta \mathcal{F}(\lambda) \right].
\end{eqnarray}
Thus, in order to achieve NNLL, corrections to both the resolved contribution ($\delta \mathcal{F}$) and the unresolved one, together with virtual corrections ($g_3$), must be computed. In Ref.~\cite{Banfi:2014sua} the calculation of $\delta \mathcal{F}$ is performed.
We remind the Reader that the NLL calculation of $\mathcal{F}$ was essentially done in the soft and collinear limit. This hypothesis must be relaxed if we are to achieve NNLL. However, the crucial observation is that we are only interested in corrections which are one power of $\as$ higher than NLL. Therefore, it is sufficient to relax the hypotheses that went into the NLL calculation for one emission at the time.
In particular, the ingredients that go into the NNLL determination of the resolved emission contributions are~\cite{Banfi:2014sua}
\begin{itemize}
\item exact rapidity bound and running coupling corrections to the
  soft and collinear function $\mathcal{F}(v)$
\item one of the emissions $k_i$ is collinear but not soft,
  generating hard-collinear and recoil corrections;
\item one of the emissions $k_i$ is soft but at wide angle;
\item gluon decay is treated non-inclusively, giving rise to a
  correlated-emission correction;
 \item two emissions are close in angle (rapidity); this contribution usually vanishes for event shapes, but does play a role in other observables, e.g.\ jet vetoes.
\end{itemize}
Correspondingly, NNLL corrections to the resummed exponent, encoded in the function $g_3$, must be computed. Although the Authors of Ref.~\cite{Banfi:2014sua} do not provide a generic computation for all these contributions, they crucially identify classes of event shapes that share the same $g_3$ function, thus enabling them to perform NNLL resummation for a fairly large class of event shapes, having as input the relatively few observables for which NNLL resummation was already known.

\section{Non-global logarithms}\label{sec:non_global}

\begin{figure}[tb]
  \begin{center}
\includegraphics[scale=0.9]{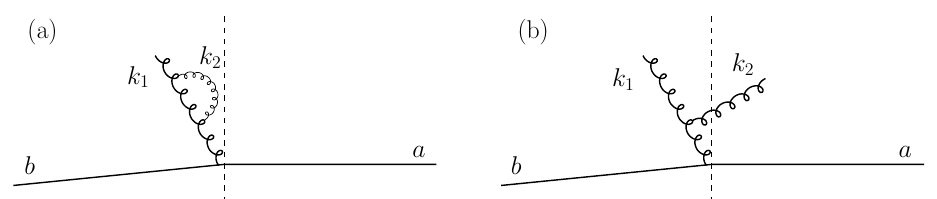}
    \caption{Kinematic configurations that give rise to non-global logarithms to lowest order in perturbation theory. Gluon $k_1\in S_1$ does not contribute to the observable, while $k_2 \in S_2$ does. This figure is adapted from Ref.~\cite{Dasgupta:2001sh}.}
    \label{fig:twogluons}
  \end{center}
\end{figure}

In the previous section we have presented all-order calculations which allow for resummation of large logarithmic corrections to NLL accuracy for a large class of event shapes. The two key properties that an observable must satisfy in order for that analysis to hold are IRC safety and globalness. In this section, we will relax the second hypothesis and we will discuss the resummation of so called non-global observables~\cite{Dasgupta:2001sh,Dasgupta:2002bw}. 
In the \caesar framework, non-globalness can be viewed as a particular failure of rIRC safety.
Not only this is an interesting theoretical question from the point of view of understanding the all-order structure of perturbative QCD, but it is also central for phenomenological studies. Many observables studied at lepton and hadron colliders are sensitive only to a restricted region of phase-space. For instance, when considering the properties of hadronic jets, we are asking questions about the energy-momentum flow within a certain phase-space region, the jet, while ignoring the outside.
We start our discussion with a fixed-order example, which will illustrate how a single logarithmic contribution arises in non-global observables, while being absent for global ones.

In the previous section we have analyzed in some detail the kinematics and the resummation of thrust. In particular, we have related the event shape $\tau$ to the sum of the invariant masses of the two hemispheres $S_1$ and $S_2$. From that analysis, we have learnt that, in order to capture NLL accuracy, it is sufficient to examine the emission of soft and collinear partons (real and virtual) as if they were emitted independently. All contributions originating from correlated emissions were accounted for by the running of the strong coupling.

Let us analyze the situation in more detail, focussing on the first non-trivial order, i.e.\ $\ord \left(\as^2\right)$. At this order we consider the emission of two gluons with momentum $k_1$ and $k_2$. If the gluons are both emitted off the quark-antiquark line their contribution is proportional to the color factor $\cf^2$. We can also encounter a situation where one gluon is emitted and then splits into a gluon pair, both ending up in the same hemisphere. This contribution is a correlated emission, proportional to $\ca \cf$, which is captured by the running of the strong coupling in the CMW scheme~\cite{Catani:1990rr}. Let us also consider the case in which the emitted gluon splits into two gluons, each of which ends up in different hemispheres. To this order we then have
$
\tau= \frac{k_1^2}{Q^2}+\frac{k_2^2}{Q^2}+\ord \left(\frac{k_i^2 k_j^2}{Q^4} \right).
$
We note that the limit in which $k_2$ is much softer than $k_1$, $k_2$ gives a vanishing contribution to thrust. Therefore, when real and virtual corrections are added together, the cancellation of the singularities is complete and no large logarithm appear from this configuration.
Thus, because the phase-space for thrust is fully inclusive we are able to conclude that correlated emissions are fully accounted for by the running of $\as$, in the CMW scheme.

What would happen if instead of thrust, i.e.\ the sum of the hemisphere masses, we were interested in the mass of one hemisphere, say $S_2$? The analysis concerning independent emissions and running coupling will go through in a similar way. Indeed, in the previous section, we built the resummation of thrust as the sum of the two jet mass contributions. However, the story dramatically changes  for configurations in which gluon $1$ is emitted in $S_1$ and gluon $2$ in $S_2$, as depicted in Fig.~\ref{fig:twogluons}. In this case $k_1$ does not contribute to the observable because it is in the wrong hemisphere $S_1$, while $k_2$ does, regardless of its softness. Cancellation of real and virtual diagrams is therefore incomplete and a large logarithm is left behind.

Let us work through this example explicitly. We need to consider the matrix element square for the emission of two soft gluon off a $q \bar q$ dipole, in the limit where $k_2$ is much softer than $k_1$~\cite{Catani:1983bz,Dokshitzer:1992ip}. This can be written as the sum of two pieces: independent and correlated emissions
\begin{equation}\label{eq:2gluons}
W= \cf^2 W^{(\rm ind)}+ \cf \ca W^{(\rm corr)},
\end{equation}
where
\begin{eqnarray}\label{eq:correlated}
W^{(\rm ind)}&=&   \frac{2\, p\cdot {\bar p}}{p\cdot k_1 \, {\bar p} \cdot k_1}   \frac{2\, p\cdot {\bar p}}{p\cdot k_2 \, {\bar p} \cdot k_2} \\
W^{(\rm corr)}&=&  \frac{2\, p\cdot {\bar p}}{p\cdot k_1 \, {\bar p} \cdot k_1}
 \left( \frac{p\cdot k_1}{p\cdot k_2 \, k_1 \cdot k_2}+ \frac{{\bar p}\cdot k_1}{{\bar p}\cdot k_2 \, k_1 \cdot k_2}-
  \frac{p\cdot {\bar p}}{p\cdot k_2 \, {\bar p} \cdot k_2}
 \right). \nonumber \\
\end{eqnarray}
In order to compute the non global contribution to the cumulative distribution $\Sigma$, we have to integrate the above matrix element over the appropriate phase-space. Adding together real and virtual contribution, we obtain
\begin{eqnarray}\label{eq:sigma_NG}
\Sigma^{(\rm ng)}_2(v)&=&- \cf \ca \left(\frac{\as}{2\pi } \right)^2 \int_{k_1 \notin S_2} [\d k_1] \int_{k_2 \in S_2} [\d k_2] \, W^{(\rm corr)}(k_1,k_2)
\nonumber \\ &\times &\Theta \left(2 \bar p \cdot k_2 - Q^2v  \right)= -\cf \ca \left( \frac{\as}{2 \pi}\right)^2 \frac{\pi^2}{3} \ln^2\frac{1}{v}.
\end{eqnarray}
Thus, the hemisphere mass distribution receives a non-global contribution starting from $\ord\left(\as^2 \right)$, which is $\as^2 L^2$, i.e.\ it contributes to NLL. We note that the logarithm in Eq.~(\ref{eq:sigma_NG}) originates from the energy integrals: non-global logarithms are related to large-angle emissions.  The coefficient of this logarithm is fixed by the angular integrations, which exhibits an integrable singularity when the gluons' momenta are close together, i.e.\ at the boundary between the two hemispheres~\cite{Dasgupta:2002bw}.

The result in Eq.~(\ref{eq:sigma_NG}) represents only the leading term at the first order at which non-global logarithms appear. In order to achieve NLL accuracy these contributions must be resummed to all-orders. As we shall discuss in the next section this is very non-trivial, even if our aim is to resum the leading tower of non-global logarithms.

\subsection{Resummation of non-global logarithms in the large $N_C$ limit} \label{sec:ng_res}
In order to perform an all-order analysis of non-global logarithms, we must consider configurations of many soft gluons. If we restrict ourselves to considering their leading contributions, which we recall is single-logarithmic, we can assume energy-ordering; however, no collinear approximation can be made. Thus, we have to describe the emission of a softer gluon off an ensemble of large-angle soft gluons. As previously discussed, color correlations make the color algebra highly non-trivial as every emission increases the dimensionality of the relevant color space. Moreover, describing the geometry of such ensemble becomes also difficult. The approach that was taken in the first analysis of non-global logarithms was to consider the large-$N_C$ limit. Color correlations becomes trivial in this limit because the off-diagonal entries of the color matrices vanish. Thus, we are able to write the matrix element square for the $n$ gluon ensemble in a factorized way~\cite{Bassetto:1984ik}
\begin{eqnarray}\label{eq:multi_gluon_soft}
W_n(p,k_1,\dots, k_n, \bar p)=\frac{1}{n!} \prod_{i=1}^n \sum_{\pi_n} \frac{p \cdot \bar p}{p \cdot k_{i_{1}} \, k_{i_{1}} \cdot k_{i_{2}}\dots  k_{i_{n}} \cdot \bar p}
\end{eqnarray}
where the sum is over the $n!$ permutations. For instance, it is easy to verify that in the large $N_C$ limit, Eq.~(\ref{eq:multi_gluon_soft}) with $n=2$ and Eq.~(\ref{eq:2gluons}) coincide. Thus, Eq.~(\ref{eq:multi_gluon_soft})  leads to a simplified physical picture because an emission off an ensemble of $n-1$ gluons (plus the two hard patrons) reduces to the emission off each of the $n$ dipoles.
When the dipole radiates a gluon, it splits into two dipoles, originating configurations which are determined by the history of the gluon branching.
This suggests to make use of a Monte Carlo implementation, which enables one to deal numerically with the second difficulty we have mentioned, namely the complicated geometry of the multi-gluon final states. This solution was first implemented in Ref.~\cite{Dasgupta:2001sh} and subsequently used in a number of phenomenological applications, e.g.~\cite{Dasgupta:2002dc,Banfi:2006gy,Banfi:2008qs,Banfi:2010pa,Dasgupta:2012hg}. We will come back to the numerical impact of non-global logarithms when we discuss the jet mass distribution in section~\ref{sec:jetmass}.

Ref.~\cite{Banfi:2002hw} instead developed a more formal treatment of non-global logarithms. Starting from Eq.~(\ref{eq:multi_gluon_soft}) the Authors of Ref.~\cite{Banfi:2002hw} were able to derive an evolution equation, henceforth the Banfi-Marchesini-Syme (BMS) equation, which, equivalently to the Monte Carlo approach, resums the leading non-global logarithm, in the large-$N_C$ limit. In order to write down the BMS equation let us consider an observable $v$, which receives contributions only from emissions within region $X$, and not from its complement $\bar X$, respectively  $S_2$ and $S_1$ in the hemisphere mass example above. For a generic pair of primary partons $a$ and $b$, not necessarily aligned with the thrust axis, the distribution of $L=-\ln v$ obeys the following equation:
\begin{eqnarray}\label{eq:BMS}
\partial_L G_{ab}(L)&=&- \int_{X}\frac{\d^2 \Omega_k}{4 \pi } \frac{p_a \cdot p_b}{p_a \cdot k \; k \cdot p_b } G_{ab}(L)  \nonumber \\ &+&
\int_{\bar{X}}\frac{\d^2 \Omega_k}{4 \pi } \frac{p_a \cdot p_b}{p_a \cdot k \; k \cdot p_b } \left[G_{ak}(L)G_{kb}(L)  -G_{ab}(L)  \right] .
\end{eqnarray}
We note that the first contribution is linear in $G$ and is only sensitive to the emission of a soft gluon off the primary dipole. This term generates a resummed exponent completely analogous to the one for global events shapes that we have discussed in the previous section. However, the striking feature of the BMS equation is its second, non-linear, contribution, which highlights the non-global nature of the evolution. Despite the fact that the BMS equation has been derived a while ago, no closed-form analytic solution is known and solutions have been determined either numerically or iteratively. For instance, the Authors of Ref.~\cite{Schwartz:2014wha} have been recently computed an iterative solution up to five loops by exploiting underlying symmetries of the equation. In Ref.~\cite{Rubin:2010fc} the expansion is calculated even one order further by means of a Monte Carlo approach. These results have also been confirmed, and extended to finite $N_C$, by means of brute force calculations of Feynman diagrams in the soft limit~\cite{Khelifa-Kerfa:2015mma}.

It has been noted~\cite{Marchesini:2003nh} that the BMS equation has the same form as the Baliksty-Kochegov (BK) equation~\cite{Balitsky:1995ub,Kovchegov:1999yj} that describes non-linear small-$x$ evolution in the saturation regime. This correspondence has been studied in detail in Refs~\cite{Avsar:2009yb,Hatta:2009nd}, where BMS and BK were related via a stereographic projection. Because a generalization of the BK equation to finite $N_C$ exists~\cite{JalilianMarian:1997gr,Iancu:2000hn}, the correspondence between non-global logarithms and small-$x$ physics was argued to hold at finite-$N_C$ and numerical solutions have been studied~\cite{Weigert:2003mm,Hatta:2013iba}. Very recently, this correspondence was indeed mathematically established~\cite{Caron-Huot:2015bja}. We shall briefly discuss this topic in section~\ref{sec:ngl_recent}.

\subsection{An alternative approach: soft-jet expansion}\label{sec:ng_res_dressed}
A different approach to the question of resumming non-global logarithms was developed in Refs.~\cite{Forshaw:2006fk,Forshaw:2008cq} and applied to a phenomenological study of jet vetoes between hard jets in Refs.~\cite{Forshaw:2009fz,Delgado:2011tp}. In that context, because color-correlations were of primary interest, the large-$N_C$ limit did not seem adequate.

In this approach, the all-order calculation was then organized differently, in terms of the number of soft (real or virtual) gluons in the region $\bar X$, where the observable has no support:
\begin{eqnarray} \label{eq:sigma_n}
\Sigma(v)= \sum_{n_k=0} \Sigma^{(n_k)}(v), \quad k \in {\bar X}.
\end{eqnarray}
 This procedure fixes the number of external partons (the soft gluons in $\bar X$ plus the original hard legs) and for each of these configurations an all-order evolution is derived. The kinematic structure of the latter is actually relatively simple, because after real-virtual cancellation, we are just left with virtual corrections in the $X$ region, which do exponentiate. For $n_k=0$ we have no gluons in $\bar X$, which corresponds to the resummation of the global component. Even if one is formally able to write all-order expressions for each of the $\Sigma^{(n_k)}$ in Eq.~(\ref{eq:sigma_n}) in terms of abstract color operators, their actual evaluation requires fixing color bases and finding representations for the color products. As already mentioned, this poses computational issues, especially because the dimensionality of the bases increases with the number of (real) gluons in $\bar X$. Therefore, only the case $n=1$ was considered in the phenomenological analyses of Refs.~\cite{Forshaw:2009fz,Delgado:2011tp}:
 \begin{eqnarray} \label{eq:omega}
 \Sigma^{(1)}(v)&=&  \frac{1}{\sigma_0}\langle M_0| \int_{k \in {\bar X}} [\d k]  \Bigg[ e^{- \bf {R^{(0)}}^\dagger}
  {\mathbf{D}^\mu}^{\dagger} \mathbf{D}_\mu e^{- \bf R^{(0)}} \nonumber \\
 &&-e^{- \bf {R^{(0)}}^\dagger}  {\mathbf{D}^\mu}^{\dagger} e^{- \bf {R^{(1)}}^\dagger} e^{- \bf R^{(1)}}\mathbf{D}_\mu  e^{- \bf R^{(0)}} \Bigg]  |M_0 \rangle,
\end{eqnarray}
where the operator $\mathbf{D}_\mu$ describes the change in color, spin and kinematics due to the emission of a soft gluon with momentum $k$.
The first line correspond to a virtual emission, which does not change the dimensionality of the color space. Therefore, the system evolves with the same resummed exponent as the global contribution $\bf {R^{(0)}}$, where the superscript counts the number of gluons in $\bar X$, zero in this case. However, in case of a real emission, as in the second line, we have now the evolution of a larger system, which is controlled by a different resummed exponent $\bf {R^{(1)}}$.

We finish this discussion pointing out that an approach similar in spirit was recently developed in the context of jet substructure using techniques of SCET~\cite{Larkoski:2015zka}, where the equivalence between the soft-jet expansion (in the large-$N_C$ limit) and an iterative solution of the BMS equation was established. Moreover, the soft-jet expansion was found to converge rapidly to the Monte Carlo solution of Ref.~\cite{Dasgupta:2001sh}.

\subsection{Recent developments}\label{sec:ngl_recent}
\begin{figure}
\begin{center}
\begin{equation*}
 \quad  \quad \quad \quad  \quad \quad \quad
   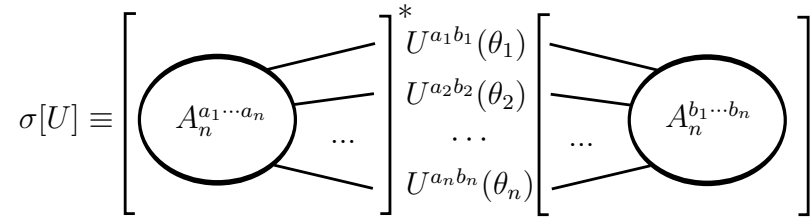
\nonumber
\end{equation*}
\caption{Pictorial representation of the color density matrix $\sigma[U]$.
For each colored final state, an independent color rotation $U^{ab}(\theta)$ is
applied between the amplitude $A_n$ and its complex conjugate. Figure adapted from Ref.~\cite{Caron-Huot:2015bja}.
}\label{fig:color-density}
\end{center}
\end{figure}
In section~\ref{sec:hadron_jets} we will discuss the growing interest in jets physics and their substructure.
This not only has lead to a rich phenomenological and experimental program in LHC physics, but it has also triggered a renaissance of all-order QCD calculations to better understand new ideas and techniques.  The long-standing issue of resumming non-global logarithms has greatly benefited from this renewed interested. A remarkable study has been recently presented in Ref.~\cite{Caron-Huot:2015bja}. In this approach, the \emph{color density matrix} $\sigma[U]$ is introduced, with the aim of describing soft radiation, as pictorially described in Fig.~\ref{fig:color-density}. An evolution equation (henceforth the Caron-Huot equation) is then derived for $\sigma[U]$, to all-loops, at finite $N_C$. The related anomalous dimension $K$ is  explicitly computed to one and two loops. The one-loop approximation to the Caron-Huot equation coincides with the BMS equation, once the large-$N_C$ limit is taken and it confirms on a firmer ground the results of Refs.~\cite{Weigert:2003mm,Hatta:2013iba} at finite $N_C$. More importantly, the explicit calculation of the two-loop contribution to $K$ paves the way for the resummation of non-global logarithms at higher-logarithmic accuracy, although computing solutions to the evolution equation remains a challenging task.

\section{Jet physics}\label{sec:hadron_jets}

Jets, i.e.\ collimated sprays of particles, are key objects in particle physics.
Jets really live at the boundary between experimental and theoretical particle physics and are abundantly used by both communities.
Indeed, the majority of physics analyses from the ATLAS and CMS collaborations uses jets as input.
Consequently, jet definitions, commonly referred to as jet algorithms, have to make sense from an experimental viewpoint as well from a theoretical one. For instance, jet algorithms should be IRC safe, so that they yield finite cross-sections when evaluated in perturbation theory~\cite{Sterman:1977wj}~\footnote{For the interested Reader, we recommend the comprehensive review on jet physics of Ref.~\cite{Salam:2009jx}.}.

Modern jet algorithms are based on the concept of sequential recombination. Pairwise distances between particles are evaluated in order to decide whether to recombine two particles. The metric used to evaluate these distances characterizes the jet algorithms. Nowadays, the most popular group of jet algorithm is the generalized $k_t$ family, for which the metric is defined by
\begin{equation}\label{eq:dij}
d_{ij} = \min\left(p_{ti}^{2p},p_{tj}^{2p} \right) \frac{\Delta R_{ij}^2}{R^2}, \quad d_{iB}=p_{ti}^{2p},
\end{equation}
where $p_{ti},p_{tj}$ are the particles' transverse momenta and $\Delta R_{ij}^2$ is their distance in the azimuth-rapidity plane. $R$ is an external parameter, which plays the role of the jet radius. Different choices for the parameter $p$ are possible. For instance, $p=0$ corresponds to the so-called Cambridge-Aachen (C/A) algorithm~\cite{Dokshitzer:1997in,Wobisch:1998wt}, with a purely geometrical distance. For $p=1$ we have the $k_t$-algorithm~\cite{Catani:1993hr,Ellis:1993tq}, which by clustering particles at low $p_t$ first, is likely to faithfully reconstruct a QCD branching history. Finally, with the choice $p=-1$ we obtain the anti-$k_t$ algorithm~\cite{Cacciari:2008gp}, which clusters soft particles around a hard core, producing fairly round jets in the azimuth-rapidity plane. It is interesting to note that all algorithms of the generalized $k_t$ family act identically on a configuration with just two particles: they are recombined if $\Delta R_{ij}<R$.

\subsection{Jet masses}\label{sec:jetmass}
Many observables have been devised to study the internal properties of jets. Some of them were originally defined to describe the properties of entire events, and then adapted to probe jets. Examples include jet shapes~\cite{Seymour:1997kj}, angularities~\cite{Berger:2003iw,Ellis:2010rwa}, and energy-energy correlation functions~\cite{Larkoski:2013eya} of high-$p_t$ jets. Perhaps the simplest example of such observables is the jet invariant mass. More precisely, we can define the dimensionless ratio
\begin{equation} \label{eq:rho}
\rho = \frac{m_{\rm jet}^2}{R^2 p_t^2}=\frac{\left(\sum_{i \in {\rm jet}} p_i\right)^2}{R^2 p_t^2}.
\end{equation}
The above ratio is small in the boosted regime $m_{\rm jet} \ll R p_t$, which is of particular interested at the LHC, as discussed later in section~\ref{sec:grooming}. Thus, in order to obtain reliable predictions for this observable, we need to perform all-order calculations.

The resummation of the $\rho$ distribution is clearly closely related to the resummation of thrust and of the hemisphere mass, which we have described in detail in the previous sections. Because large-angle radiation contributes to single-logarithmic accuracy, in order to achieve NLL, we have to take into account soft gluons emitted from each of the dipoles. However, in order to make our discussion simpler, we can work in the small-$R$ limit, neglecting contributions that vanish as powers of $R$. In this limit a simple picture emerges because large-angle radiation from dipoles other than the one involving the measured jet is suppressed~\cite{Banfi:2010pa}. Corrections to this picture can be systematically included~\cite{Dasgupta:2012hg} as a power series in the jet radius $R$. Moreover, logarithmic contributions at small-$R$ can also be resummed~\cite{Dasgupta:2014yra}. Thus, the  NLL cumulative distribution for an isolated jet reads
\begin{eqnarray}\label{eq:jet_mass_res}
\Sigma(\rho) &=& \frac{\exp \left[\frac{1}{2\as} g_1(\lambda_\rho) +\frac{1}{2}g_2(\lambda_\rho) \right] }{\Gamma \left(1-\left(2\as\right)^{-1}\partial_L g_1(\lambda_\rho)\right)} \,
\Sigma^{(\rm ng)}(\lambda_\rho) \,
\Sigma^{(\rm alg)}(\lambda_\rho),
\end{eqnarray}
where in case of a quark jets the functions $g_i$ are defined in Eq.~(\ref{eq:g1_g2}) and the factor of $1/2$ appears because thrust essentially corresponds to sum of the two hemisphere masses.

We have already discussed the non-global factor $\Sigma^{(\rm ng)}$ and its resummation in section~\ref{sec:non_global}. Here we limit ourselves to note that non-global logarithms do not vanish in the small-$R$ limit.
Moreover, the detailed form of the non-global contributions also depends on the clustering algorithm that defines the jet. In particular, in the presence of many soft emissions together with a hard parton, the anti-$k_t$ algorithm will always cluster all soft gluons to the hard parton, behaving as a rigid cone algorithm. In this case the algorithm provides a sharp boundary and non-global logarithms turn out to be the same as in the hemisphere case, up to corrections suppressed by powers of the jet radius.

 The function $\Sigma^{(\rm alg)}$ accounts for departure from the rigid-cone situation and therefore $\Sigma^{({\rm anti-}k_t)}=1$. The choice of different algorithms, such as C/A or $k_t$ can result into a reduction of the non-global contribution~\cite{Appleby:2002ke}, however these algorithms are characterized by a non-trivial $\Sigma^{(\rm alg)}$~\cite{Banfi:2005gj,Delenda:2006nf,Banfi:2010pa}.

\begin{figure}
\begin{center}
\includegraphics[scale=0.6]{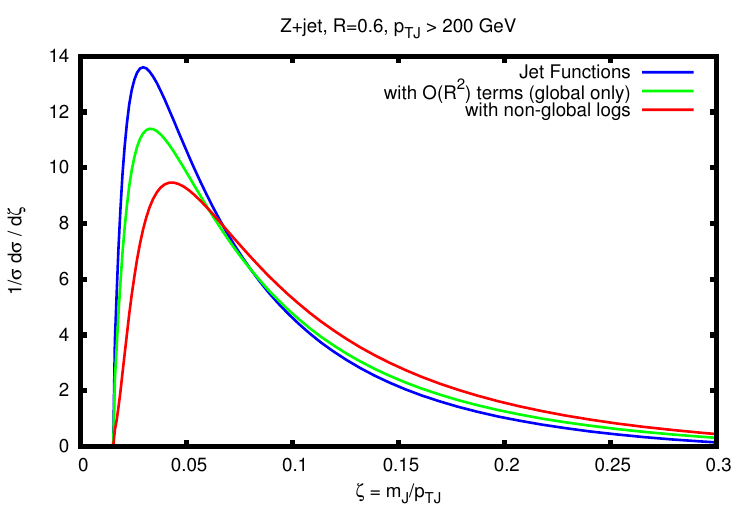}
\includegraphics[scale=0.6]{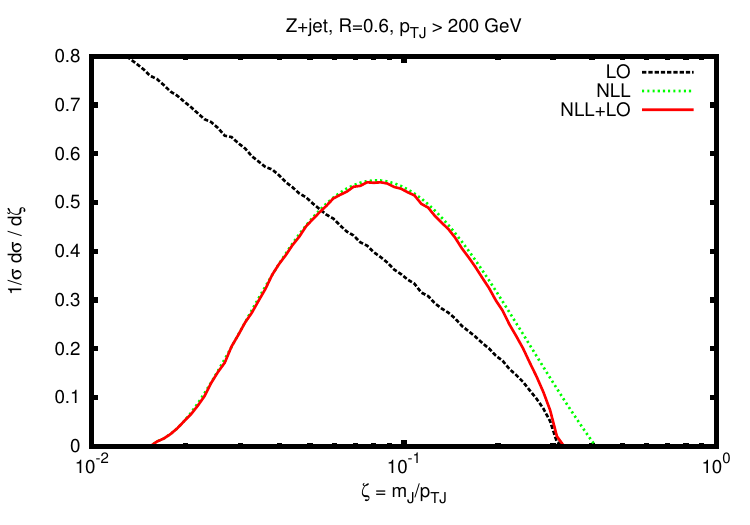}
\end{center}
\caption{The jet mass distribution of the hardest jet in associated production of a $Z$ boson with (at least) one jet, plotted as a function of $\zeta = R \sqrt{\rho}$. The plot on the left shows the NLL resummation in three different approximations: in the small-$R$ limit (blue curve, labelled as ``Jet Functions''), with finite $R^2$ corrections (in green) and with non-global logarithms (in red).
The plot on the right shows the fixed-order $\ord{\left(\as\right)}$ calculation (dotted black curve), the NLL resummation (green dotted curve) and the matching of the two (solid red curve).
Both figures are taken from Ref.~\cite{Dasgupta:2012hg}.}
\label{fig:jetmass}
\end{figure}

The jet mass distribution of the hardest jet in associated production of a $Z$ boson with at least one jet, plotted as a function of $\zeta = R \sqrt{\rho}$ is shown in Fig.~\ref{fig:jetmass}. Jets are defined using the anti-$k_t$ algorithm.

The plot in Fig.~\ref{fig:jetmass}, on the left, shows the NLL resummation in three different approximations.
The result in the isolated jet approximation of Eq.~(\ref{eq:jet_mass_res}) is plotted in blue. Corrections due to the finite size of the jet radius are accounted for in the green curve. We note that all dipoles (three in this case) contribute now and the bulk of these corrections come from initial-state radiation. Finally, the resummation of non-global logarithms is accounted for in the red curve.

The plot in Fig.~\ref{fig:jetmass}, on the right, shows instead a comparison of the fixed-order $\ord(\as)$ result (black), the resummed one (green) and the matching between the two (red). At small values of $\zeta$, the fixed-order curve exhibits the expected logarithmic divergence and in order to obtain a reliable prediction we have to resum these contributions to all orders. However, the resummed result is based on the eikonal approximation which is not accurate when $m_{\rm jet}\sim R p_t$. We can obtain a reliable description of the entire spectrum by matching the resummation to the fixed-order result, as shown in red.

\subsection{Grooming and tagging algorithms}\label{sec:grooming}
Because of its unprecedentedly high colliding energy, the LHC is reaching energies far above the electro-weak scale. Therefore, analyses and searching strategies developed for earlier colliders, in which electro-weak scale particles were produced with small velocity, have to be fundamentally reconsidered. In particular, in the context of jet-related studies, the large boost of electro-weak objects (not only Standard Model particles like electro-weak and Higgs bosons and top quarks, but also any new particle with a mass of the order of the electro-weak scale) causes their hadronic decays to become collimated inside a single big jet~\cite{Seymour:1993mx,Butterworth:2002tt}.

This is particularly important in the context of Higgs physics, because its dominant decay channel is into $b$-jets, which suffers from a huge QCD background. However, when the Higgs is produced with large transverse momentum, its decay products are likely to be reconstructed in one big jet. The presence of the Higgs can be then inferred by studying the substructure of this jet~\cite{Butterworth:2008iy}.
Consequently, jet substructure has emerged as an important tool for searches at the LHC and a vibrant field of theoretical and experimental research has developed in the past few years, producing a variety of studies and techniques~\cite{Abdesselam:2010pt,Altheimer:2012mn,Altheimer:2013yza,Adams:2015hiv}.

Many ``grooming" and ``tagging" algorithms, e.g.\ the mass-drop tagger~\cite{Butterworth:2008iy}, trimming~\cite{Krohn:2009th}, pruning~\cite{Ellis:2009su,Ellis:2009me}, have been developed, successfully tested and are currently used in experimental analyses. Broadly speaking, a grooming procedure takes a jet as an input and tries to clean it up by removing constituents which being at wide angle and relatively soft, are likely to come from contamination, such as the underlying event or pile-up. A tagging procedure instead focuses on some kinematical variable that is able to distinguish signal from background, such as, for instance, the energy sharing between two prongs within the jet, and cuts on it. Many of the algorithms on the market usually perform both grooming and tagging and a clear distinction between the two is difficult.

Regardless of their nature, substructure algorithms try to resolve jets on smaller angular and energy scales, thereby introducing new parameters. This challenges our ability of computing predictions in perturbative QCD.
Indeed, most of the theoretical studies of substructure tools have been done using Monte Carlo parton showers. While these are powerful general purpose tools, their essentially numerical nature offers little insight into the results produced or their detailed and precise dependence on algorithm parameters. In the past few years however, we have reached a deeper understanding of, at least, the most basic (and most used) grooming and tagging techniques, both in the presence of background~\cite{Dasgupta:2013ihk,Dasgupta:2013via} and signal jets~\cite{Feige:2012vc,Dasgupta:2015yua}. This understanding has been put at work and new substructure algorithms, which combine efficient signal / background discrimination together with robust theoretical understanding, have been devised. One of them is soft-drop~\cite{Larkoski:2014wba}, which we will use as an example for our discussion.

The soft-drop procedure takes a C/A jet and implements the following steps:
\begin{enumerate}
\item Break the jet $j$ into two subjets by undoing the last stage of C/A clustering.  Label the resulting two subjets as $j_1$ and $j_2$.
\item \label{step:grooming} If the subjets pass the soft drop
  condition $\frac{\min(p_{t1},p_{t2})}{p_{t1}+p_{t2}}>\zcut
  \left(\frac{\Delta R_{12}}{R} \right)^\beta$, then deem $j$ to be the final soft-drop
  jet.
\item Otherwise, redefine $j$ to be equal to subjet with larger $p_t$ and iterate the procedure.
\item If $j$ is a singleton and can no longer be declustered, then one can either remove $j$ from consideration (``tagging mode'') or leave $j$ as the final soft-drop jet (``grooming mode'').
\end{enumerate}

\begin{figure}
\begin{center}
\includegraphics[scale=0.6]{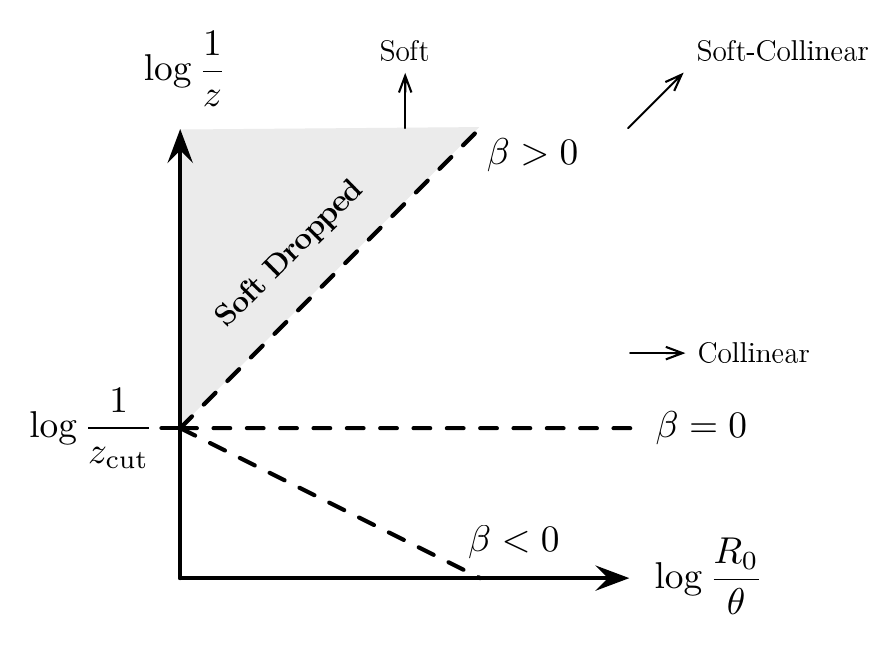}
\end{center}
\caption{Phase space for emissions on the $(\ln \frac{1}{z},\ln \frac{R}{\theta})$ plane. For $\beta > 0$, soft emissions are vetoed while much of the soft-collinear region is maintained.  For $\beta = 0$, both soft and soft-collinear emissions are vetoed.  For $\beta < 0$, all (two-prong) singularities are regulated by the soft drop procedure. Figure taken from Ref.~\cite{Larkoski:2014wba}.}
\label{fig:groomedregions}
\end{figure}

The difficulty posed by substructure algorithms in general, and soft drop in particular, is the presence of new parameters that slice the phase-space for soft gluon emission in a non-trivial way, resulting in potentially complicated all-order behavior of the observable at hand.
This is exemplified in Fig.~\ref{fig:groomedregions}, where we show the phase space for a single gluon emission from a  hard quark/gluon in the $(\ln \frac{1}{z},\ln \frac{R}{\theta})$ plane, where $0 \le z \le 1$ is the energy fraction of the emitted gluon with respect to the hard parton initiating the jet and $0 \le \theta \le R$ is the angle of the emission, measured from the hard parton.
This figure is analogue to the one we have used Fig.~\ref{fig:phase-space} to describe the allowed phase-space for event shapes, but now expressed in $(\ln \frac{1}{z},\ln \frac{R}{\theta})$ variables rather than $(\ln \frac{Q}{k_t},\eta)$. In either set of coordinates the emission probability is flat in the soft-collinear limit.

In the soft limit, the soft drop criterium reduces to
\begin{equation}
\label{eq:vetoline}
z > \zcut \left(\frac{\theta}{R}\right)^\beta \quad \Rightarrow \quad \ln \frac{1}{z} < \ln \frac{1}{\zcut} + \beta \ln \frac{R}{\theta}.
\end{equation}
Thus, vetoed emissions lie above a straight line of slope $\beta$ on the $(\ln \frac{1}{z},\ln \frac{R}{\theta})$ plane, as shown in Fig.~\ref{fig:groomedregions}.
For $\beta > 0$, collinear splittings always satisfy the soft drop condition, so a soft-drop jet still contains all of its collinear radiation.  The amount of soft-collinear radiation that satisfies the soft drop condition depends on the relative scaling of the energy fraction $z$ to the angle $\theta$.  As $\beta\to 0$, more of the soft-collinear radiation of the jet is removed, and in the $\beta=0$ limit, all soft-collinear radiation is removed~\footnote{Soft-drop with $\beta=0$ corresponds to the modified Mass Drop Tagger~\cite{Dasgupta:2013ihk,Dasgupta:2013via}.}.  Therefore, we expect the coefficient of the double logarithms in observables like groomed jet mass, the origin of which is soft-collinear radiation,  to be proportional to $\beta$.
In the strict $\beta = 0$ limit, collinear radiation is only maintained if $z > \zcut$.  Because soft-collinear radiation is vetoed, the resulting jet mass distributions will only exhibit single logarithms, as emphasized in~\cite{Dasgupta:2013ihk,Dasgupta:2013via}.
Moreover, non-global logarithms were found to be power-suppressed for $\beta>0$, and absent for $\beta=0$.
Finally, for $\beta < 0$, there are no logarithmic structures for observables like groomed jet mass at arbitrarily low values of the observable.  For example, $\beta = -1$ roughly corresponds to a cut on the relative transverse momentum of the two prongs under scrutiny.

\begin{figure}
\begin{center}
\includegraphics[scale=0.8]{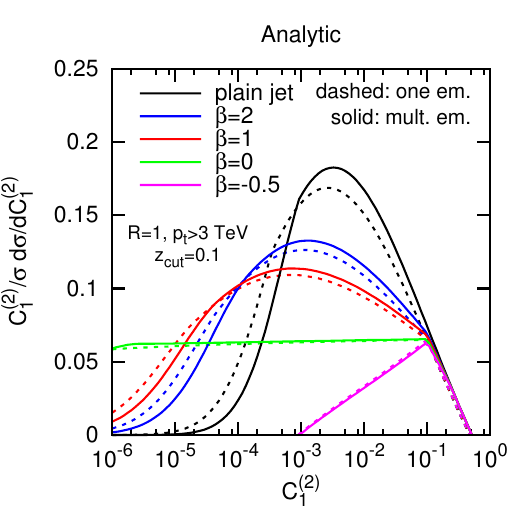}
\includegraphics[scale=0.8]{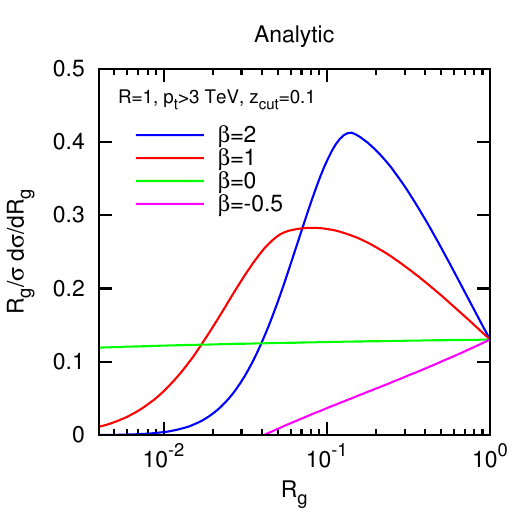}
\end{center}
\caption{Two examples of NLL resummation for groomed observables. The grooming procedure of choice is soft-drop~\cite{Larkoski:2014wba}. The plot on the left shows the energy-correlation function~\cite{Larkoski:2013eya} after soft drop, for different values of $\beta$. The plot on the right shows the groomed jet radius, i.e. the distance in azimuth and rapidity of the first pair of prongs that passes soft drop. Both plots are taken from Ref.~\cite{Larkoski:2014wba}.}
\label{fig:groomedstuff}
\end{figure}

The above understanding can be formalized into actual calculations and the resummation of a variety of observables measured on soft-drop jets has been performed to NLL~\cite{Larkoski:2014wba}. Examples of such observables include the jet mass, energy-correlation functions, the effective jet radius after grooming, the energy removed by the grooming procedure, as well as the energy sharing of first splitting that passes soft drop~\cite{Larkoski:2014bia,Larkoski:2015lea}. Two examples are shown in Fig.~\ref{fig:groomedstuff}: on the left the groomed energy-correlation function, and on the right the groomed jet radius, that is the distance in azimuth and rapidity of the first pair of prongs that passes soft drop. The different curves are for different values of the soft-drop parameter $\beta$, from no grooming (i.e. $\beta=\infty$), down to very aggressive grooming (negative $\beta$).

One interesting feature that can be observed is the presence of a transition point in the distributions. Its position can be analytically predicted as a function of the algorithm's parameters and jet's kinematics.
Moreover, the logarithmic structure of the distribution changes when we move across this transition point. In order to understand both of these features is actually enough to perform a very simple calculation. We consider the $\ord\left(\as\right)$ contribution to differential distribution for the energy-correlation $\ea$~\cite{Larkoski:2013eya} in both soft and collinear limits, i.e. $z\ll1$ and $\theta \ll R$:
\begin{eqnarray}\label{ang-LO-start}
\fl
\frac{1}{\sigma}\frac{\d \sigma^{ \rm LO}}{\d \ea}=\frac{ 2 \as C_i}{\pi} \int_0^{R} \frac{\d \theta}{\theta} \int_0^1 \frac{\d z}{z}
\, \Theta\left(z-\zcut \left(\frac{\theta}{R} \right)^\beta \right) \, \delta \left(\ea - z \left(\frac{\theta}{R} \right)^\alpha \right),
\end{eqnarray}
where $C_i=\cf (\ca)$ in case of a quark (gluon) initiated jet. The $\delta$ function fixes a measured value of $\ea$, in the soft-collinear limit, while the $\Theta$ function enforces the emission to pass soft drop, see Eq.~(\ref{eq:vetoline}).

For $\beta \ge 0$, the evaluation of the two integrals is straightforward:
\begin{eqnarray}
\label{ang-LO-res}
 \frac{\ea}{\sigma}\frac{\d \sigma^{\rm LO}}{\d \ea} &\simeq& \frac{2\as C_i}{\pi}\frac{1}{\alpha} \Bigg[
\ln \frac{1}{\ea} \Theta \left( \ea -z_{\rm cut} \right) \nonumber \\&&+
\frac{\beta}{\alpha+\beta} \ln \frac{1}{\ea}+ \frac{\alpha}{\alpha+\beta}\ln \frac{1}{z_{\rm cut}} \Theta \left(  z_{\rm cut}-\ea \right) \Bigg],
\end{eqnarray}
while for $\beta < 0$, there is an additional restriction which imposes a minimum allowed value for the observable $ \ea> {z_{\rm cut}}^{\alpha/ |\beta|}$.
Thus, we see that the result has indeed a transition point: for $\ea > z_{\rm cut}$ the distribution is insensitive to the grooming procedure, while soft drop becomes active for $\ea < z_{\rm cut}$. Moreover, the coefficient of the double logarithm in this region is proportional to $\beta$ and consequently vanishes in the $\beta \to 0$ limit, as previously anticipated. It is also interesting to note that a direct $\ord\left(\as^2\right)$ shows that non-global logarithmic contributions to the $\ea$ spectrum are actually power-suppressed in the soft drop region $\ea < z_{\rm cut}$.

The results discussed in this section, demonstrate that our understanding of the field of jet substructure has reach a level of maturity comparable to what has been achieved for more traditional observables, like the ones we have previously been discussing in this review.

\section{An unsettling end: breakdown of factorization} \label{sec:factorization}

The concept of factorization, i.e.\ the ability of separating physical effects that happens at different energy scale is the foundation of all the resummation program we have been discussing so far. Even more generally, we can say any QCD calculation, being it done at fixed-order or at the resummed level requires some notion of factorization.
 Of particular importance is the collinear factorization theorem~\cite{Collins:1983ju,Collins:1985ue,Collins:1989gx,Collins:1988ig,Collins:1998ps} that allows us to separate the perturbative, i.e.\ calculable, part of a process from the non-perturbative one, which can be described in terms of parton distribution (or fragmentation) functions. These objects are universal, i.e.\ they do not depend on the particular process, and can be determined by fitting data from previous experiments.

Although explicitly proven only for a very few inclusive process, such as, for instance deep inelastic scattering of an electron off a proton~\cite{Bardeen:1978yd}, the Drell-Yan invariant mass and rapidity distribution~\cite{Bodwin:1984hc, Collins:1983ju,Collins:1985ue}, and heavy quarkonium production~\cite{Nayak:2005rt}, collinear factorization is usually considered valid and is used ubiquitously in perturbative QCD calculations, regardless of their inclusive or exclusive nature.
 For instance, Monte Carlo parton showers are built on the idea of all-order factorization.

The Authors of Ref.~\cite{Catani:2011st} embarked in an analysis of factorization properties of partonic scattering amplitudes and their related cross-sections. The analysis shows that strict collinear factorization of QCD amplitudes is violated  beyond tree-level for initial-state (space-like) splittings. These factorization-breaking contributions originate from the exchange of Coulomb modes between the two incoming partons long before the hard interaction or between two outgoing partons, much later than the hard interaction~\cite{Forshaw:2012bi}. As we have already seen, this results into an imaginary part, which cancels when evaluating the one-loop squared amplitude. However, as mentioned in section~\ref{sec:caesar}, this cancellation is not guaranteed at subsequent orders if the color-structure of the process is non-trivial.
In particular, for the case of $2\to 2$ scattering of QCD massless partons, factorization-breaking contributions are found to cancel at two-loops, but they appear to give a non-vanishing contribution to the squared amplitude at three-loops and beyond.

Let us then discuss what happens if we attempt to perform an all-order calculation of  the (cumulative) distribution of an IRC safe observable $v$. Because we sum over all modes that live at a scale below $v$ fully inclusively, the actual singularities cancel between real and virtual corrections, ultimately because the colliding hadrons are colorless states. However, because we are measuring $v$, we have to veto real emissions above that scale. This, as discussed at length, induces potentially large logarithms, and prevents also potential factorization-breaking effects to cancel.
The consequence is remarkable: the familiar picture of a hadron-hadron cross-section made of a hard scattering coefficient function and two universal PDFs, which are evolved independently up to the hard scale, breaks down (at high enough perturbative orders) if we insist on making the cross-section more exclusive by measuring an observable at a lower scale.

The effect discussed above can have spectacular consequences on specific observables. Let us consider, for instance, dijet production in hadron-hadron collisions. We identify the two leading jets and veto the presence of a third jet in the region (in azimuth and rapidity) between the leading ones. The cross-section for this observable is affected by large logarithms of the veto scale over the hard jet transverse momentum, which we wish to resum. The Authors of Refs.~\cite{Forshaw:2006fk,Forshaw:2008cq} were set to tackle this project when they found an explicit example of factorization breaking.
According to the standard analysis performed so far, cross-sections with central jet vetoes are non-global, because phase-space is restricted only between the hard jets. The cross-section is expected to be single-logarithmic, i.e.\ $\as^n L^n$, with all the logarithms originating from soft emission away from the jets. However, a \emph{super-leading} contribution $\as^n L^{2n-3}$ is found for $n\ge 4$, when the transverse momentum of the emission is chosen as the ordering variable~\cite{Forshaw:2006fk,Forshaw:2008cq}. The extra logarithms originate from the fact that emissions outside the region between the leading jets can go forward, i.e.\ collinear to the initial-state. However, these collinear contributions are prevented to cancel against the correspondent virtual contributions precisely because of the presence of Coulomb exchanges.

Finally, we mention that the effect of factorization-breaking on global event shapes was discussed in Ref.~\cite{Banfi:2010xy}, where it was noted that the logarithmic accuracy at which these effects kick in depends on the parameters $b_l$ (and possibly $a$) and crucially on the choice of the ordering variable. If transverse momentum was confirmed to be the correct choice, then the effect would be very suppressed for $b_l \le 0$ but could potentially be relevant, even at NLL, for $b_l>0$.

\section{Summary and Conclusions}\label{sec:conclusions}

In this topical review we have discussed the basic concepts behind all-order calculations in QCD. We have focussed on calculations that make use of resummation in the so-called direct QCD framework. Similar results can also be obtained using the methods of Soft-Collinear Effective Theory.

In the first part of this review, we have discussed basic properties of gauge-theory amplitudes in the soft and collinear limits, specifically QED and QCD.
We have started by reviewing in section~\ref{sec:ir_divergences} factorization properties at one loop, and then we have generalized this analysis to all orders in section~\ref{sec:resummation}.
In the second part of this review, we have instead focussed on phenomenological applications. In section~\ref{sec:event_shapes}, we have applied the all-order techniques previously presented to the event-shape variable
thrust, discussing its NLL resummation in some detail. We have then presented the \caesar framework, which enables one to perform the resummation of a large class of event shapes and briefly mentioned its extension to NNLL. Section~\ref{sec:non_global} was instead devoted to a discussion of non-global observables and their resummation.
In section~\ref{sec:hadron_jets}, we have illustrated how logarithmic resummation is an invaluable tool to describe the physics of hadronic jets and, in particular, their internal structure, which is a topic of the highest importance in the context of new physics searches at high-energy colliders.  Finally, we have briefly discussed important issues about the limitations of the framework we have used to factorize amplitudes and cross-sections.

With this review, we hope we have managed to stimulate the curiosity of a Reader who begins to study this field, explaining them the concepts and the methods on which all-order calculations are based, with a constant eye on their phenomenological implications, as well as recent developments in the field.
We realize that a full account of technical details would require a text-book, rather than a review.
Therefore, we have put an effort in compiling an exhaustive bibliography and we encourage the Reader to read the original papers in order to gain a deeper, more technical understanding.
Because we feel that a concise summary of this kind of calculations and their applications was lacking, we also reckon this effort may prove useful also for a more expert Reader.

\ack

We thank our colleagues and collaborators Andrea Banfi, Stefano Catani, Mrinal Dasgupta, Jeff Forshaw, Andrew Larkoski, Pier Francesco Monni, Duff Neill, Gavin Salam, Mike Seymour, Gregory Soyez, George Sterman, and Jesse Thaler for many useful discussions and comments on this manuscript.
We also thank Stefano Forte for encouraging us to undertake this work.

The work of G.~L. is supported by the Alexander von Humboldt Foundation, in the framework of the Sofja Kovaleskaja Award 2010, endowed by the German Federal Ministry of Education and Research.
The work of S.~M. is supported by the U.S.\ National Science Foundation, under grant PHY--0969510, the LHC Theory Initiative.
\\
\bibliographystyle{hieeetr}

\bibliography{references}

\end{document}